\documentclass{aa}

\usepackage{graphicx}
\usepackage{txfonts}
\usepackage{footmisc}

\newcommand{\mf}{{\tt molecfit}}
\newcommand{\Mf}{{\tt Molecfit}}

\newcommand{\xshoot}{\mbox{X-Shooter}}

\begin{document}

   \title{{\tt Molecfit}: A general tool for telluric absorption correction\thanks{\tt http://www.eso.org/sci/software/pipelines/skytools/}}
   \subtitle{II. Quantitative evaluation on ESO-VLT/\xshoot\;spectra}
   \author{W. Kausch\inst{1,2}
          \and
          S. Noll\inst{1}
          \and
          A. Smette\inst{3}
          \and
          S. Kimeswenger\inst{4,1}
          \and
          M. Barden\inst{5}
          \and
          C. Szyszka\inst{1}
          \and
          A. M. Jones\inst{1}
          \and
          H. Sana\inst{6,2}
          \and
          H. Horst\inst{7}
          \and
          F. Kerber\inst{8}
          }

   \institute{Institute for Astro- and Particle Physics, University of Innsbruck,
              Technikerstr. 25/8, A-6020 Innsbruck, Austria\\
              \email{wolfgang.kausch@uibk.ac.at}
        \and
            University of Vienna, Department of Astrophysics,
            Türkenschanzstr. 17 (Sternwarte),
            A-1170 Vienna, Austria
        \and
            European Southern Observatory,
            Alonso de C\'ordova 3107, 19001 Casilla,
            Santiago 19, Chile
        \and
            Instituto de Astronom\'ia, Universidad Cat\'olica del Norte,
            Avenida Angamos 0610,
            Antofagasta, Chile
        \and
            International Graduate School of Science and Engineering,
            Technische Universit\"at M\"unchen,
            Boltzmannstr. 17,
            85748 Garching bei M\"unchen, Germany
         \and
            Space Telescope Science Institute,
            3700 San Martin Dr, Baltimore, MD 21218,
            United States
         \and
            Josef-Führer-Straße 33, 80997 M\"unchen, Germany
        \and
            European Southern Observatory,
            Karl-Schwarzschild-Str. 2,
            85748 Garching bei M\"unchen, Germany
             }

   \date{Received ; accepted }


  \abstract
   {Absorption by molecules in the Earth's atmosphere strongly affects ground-based astronomical observations. The resulting absorption line strength and shape depend on the highly variable physical state of the atmosphere, i.e. pressure, temperature, and mixing ratio of the different molecules involved. Usually, supplementary observations of so-called telluric standard stars (TSS) are needed to correct for this effect, which is expensive in terms of telescope time. We have developed the software package {\tt molecfit} to provide synthetic transmission spectra based on parameters obtained by fitting narrow ranges of the observed spectra of scientific objects. These spectra are calculated by means of the radiative transfer code LBLRTM and an atmospheric model. In this way, the telluric absorption correction for suitable objects can be performed without any additional calibration observations of TSS.}
   {We evaluate the quality of the telluric absorption correction using {\tt molecfit} with a set of archival ESO-VLT/\xshoot{} visible and near-infrared spectra. }
   {Thanks to the wavelength coverage from the $U$ to the $K$ band, \xshoot{} is well suited to investigate the quality of the telluric absorption correction with respect to the observing conditions, the instrumental set-up, input parameters of the code, the signal-to-noise of the input spectrum, and the atmospheric profiles. These investigations are based on two figures of merit, $I_\mathrm{off}$ and $I_\mathrm{res}$, that describe the systematic offsets and the remaining small-scale residuals of the corrections. We also compare the quality of the telluric absorption correction achieved with \mf{} to the classical method based on a telluric standard star.}
   {The evaluation of the telluric correction with \mf{} shows a convincing removal of atmospheric absorption features. The comparison with the classical method reveals that \mf{} performs better because it is not prone to the bad continuum reconstruction, noise, and intrinsic spectral features introduced by the telluric standard star.}
   {Fitted synthetic transmission spectra are an excellent alternative to the correction based on telluric standard stars. Moreover, \mf{} offers wide flexibility for adaption to various instruments and observing sites.}


   \maketitle

\section{Introduction}\label{sec:intro}
Ground-based observations are naturally affected by various physical
processes in the Earth's atmosphere, in particular scattering and absorption.
The dynamics of the weather conditions, seasonal effects,
or climate change lead to variabilities in temperature, pressure, and the
chemical composition on time scales from seconds to decades, making the absorption correction a demanding matter. Thus, any data calibration usually needs supplementary calibration frames. However, this approach is very expensive in terms of telescope time, because these data should be taken directly before or after the science target.

This applies particularly to the correction arising from molecular
absorption in the Earth's atmosphere. The required supplementary calibration
frames are observations of telluric standard stars (hereafter TSS). The TSS are usually bright hot stars (e.g. B-type), that show only a few, well known intrinsic spectral features. The TSS have to be observed at a similar airmass, time, and line of sight as the science target to probe the same atmosphere column.

Within the framework of the in-kind contribution of Austria's accession to the European Southern Observatory (ESO), we developed a comprehensive sky model\footnote{\url{http://www.eso.org/observing/etc/skycalc/skycalc.htm}}
covering a wavelength range from $0.3$ to $30~\mu$m \citep{NOL12}. It was originally designed for the ESO Exposure Time Calculator\footnote{\url{http://www.eso.org/observing/etc/}} and incorporates
several components, such as airglow, scattered moonlight \citep{JON13}, zodiacal light \citep{LEI98}, scattered starlight, and the telescope emission modelled as a grey body. It also contains a spectral model of the Earth's lower atmosphere in local thermal equilibrium calculated by means of the radiative transfer code LNFL/LBLRTM\footnote{\url{http://rtweb.aer.com/lblrtm_frame.html}}
\citep{CLO05}. This code is used with the spectral line parameter database {\tt aer}, which is based on HITRAN\footnote{\url{http://www.cfa.harvard.edu/hitran/}} \citep{ROT09} and delivered together with the code. Averaged atmospheric profiles for Cerro Paranal describing the chemical composition as a
function of height (combination of the ESO MeteoMonitor\footnote{\tt http://archive.eso.org/asm/ambient-server}, a standard atmosphere\footnote{\url{http://www.atm.ox.ac.uk/RFM/atm/}} and the 3D Global Data Assimilation System (GDAS) model\footnote{\tt http://www.ready.noaa.gov/gdas1.php\label{fn:gdas1}}) are also used as input.
The GDAS model is provided by the National Oceanic and Atmospheric Administration (NOAA)\footref{fn:gdas1} and contains time-dependent profiles of the temperature, pressure, and humidity.

The ESO In-Kind software also includes tools for removing the atmospheric signature in spectra\footnote{\url{http://www.eso.org/pipelines/skytools/}}$^,$\footnote{\url{http://www.uibk.ac.at/eso/software/}}. The package \mf{} is an alternative approach for performing the telluric absorption correction by means of theoretically calculated transmission spectra based on the atmospheric model. We use the previously mentioned radiative transfer code and line database to derive synthetic transmission curves. The required atmospheric input profile is created in the same way as for the sky model, but for the time and location of the observations and not averaged. The algorithm and package functionalities are described in more detail in \cite{SME14} (hereafter Paper~I). Here, we evaluate the quality of {\tt molecfit} as a telluric absorption correction tool for ESO Very Large Telescope (VLT) instruments. The focus lies on observations taken with the \xshoot{} instrument \citep{VER11}, an echelle spectrograph covering simultaneously the wavelength regime from the $U$ to the $K$ band at medium resolution. We use two methods: (a) a statistical study of the quality of the correction by means of figures of merit, and (b) a comparison of the molecfit correction with that obtained using the traditional method based on observations of TSS.

We first briefly describe the method of the telluric absorption correction incorporated in \mf{} (Sect.~\ref{subsec:tellcorr}). A description of the test data set is given in Sect.~\ref{sec:data}. In the following, we perform two different tests: In Sect.~\ref{sec:results}, we investigate the dependence of the telluric absorption correction quality on several parameters, i.e. the influence of the line transmission, observing conditions, spectral resolution, fitting ranges, input parameters, the atmospheric profiles, and the signal-to-noise ratio $S/N$. For this purpose, we introduce two figures of merit to estimate the quality of the telluric correction $I_\mathrm{off}$ and $I_\mathrm{res}$, describing the systematic offsets and the remaining small-scale residuals of the corrections, respectively. These quantities are determined for a large data set of X-Shooter spectra, to which \mf{} is applied with default parameters optimised for this instrument. The results of this analysis allow the user to estimate the achievable quality for the telluric absorption correction in case an individual parameter set optimisation is not possible, e.g. due to the large number of spectra. In Sect.~\ref{sec:comparison}, we compare the classical method based on a TSS with \mf{} as another test. We have selected four science observations, for which both methods are optimised. In this way, individually optimised results can be compared for both methods. The results of this comparison are demonstrated by focusing on the scientifically relevant spectral details. Section~\ref{sec:summary} provides a summary of our findings.


\section{Telluric correction with \mf{}}\label{subsec:tellcorr}
The correction with the \mf{} package is performed in two steps:
\begin{itemize}
\item Determination of the best-fit atmospheric and instrumental parameters: they are related to the total column densities of the input atmospheric profile and the instrumental parameters (spectral line profile, wavelength calibration, continuum position) are optimised by means of a $\chi^2$-Levenberg-Marquardt minimisation algorithm (see Paper~I for a comprehensive description of the underlying algorithms) to best reproduce the observed telluric spectrum in selected wavelength regions.
    By varying the scaling factor of the molecular profiles of the initial input atmospheric profile, the programme iteratively calculates transmission curves, which are fitted to the input science spectrum. To minimise the calculation time, to optimise the continuum fit, and/or to avoid regions affected by stellar spectral features or instrument defects, the fitting is restricted to user-defined spectral ranges.
\item Correction of the telluric spectrum: the best-fit atmospheric and instrumental parameters are used to calculate the atmospheric transmission spectrum over the entire wavelength range of the scientific observation. The science spectrum is then divided by this transmission curve to produce telluric corrected data.
\end{itemize}
\begin{table}
\caption[]{Table of objects used for detailed tests.}
\label{tab:tac_comp_dataset1}
\centering
\begin{tabular}{l l r l l}
\hline\hline
\noalign{\smallskip}
\#$^{a}$ & object  &  median counts & slit & Prog.-ID\\
   & type    &   [ADU]       & [$\arcsec\times\arcsec$] & \\
\noalign{\smallskip}
\hline
\noalign{\smallskip}
1 & Seyfert 2 & 462  & 1.5x11 & 087.B-0614\\
2 & GRB           &    241  & 0.9x11$^{b}$ & 091.C-0934\\
\noalign{\smallskip}
\hline
\noalign{\smallskip}
3  & B[e] star    &    7684 & 0.4x11 & 084.C-0952\\
4  & E0 galaxy    &    5236 & 0.9x11 & 384.B-1029\\
5  & Carbon star  &   50587 & 0.9x11 & 084.D-0117\\
6 & PN            &   6214  & 0.4x11  & 385.C-0720\\
\noalign{\smallskip}
\hline
\end{tabular}
\tablefoot{\\
\tablefoottext{a}{Objects \#1 and \#2 are data with low $S/N$ used for tests on the reliability of \mf{} when applied to such data. \#3 through \#6 are used for a detailed comparison with the classical method. \\}
\tablefoottext{b}{observed with the $K$ band blocking filter \citep{VER11}}
}
\end{table}
\section{The data set}\label{sec:data}
In order to evaluate the performance of \mf{}, we have used archival data obtained with the \xshoot{} instrument mounted at the ESO VLT \citep{VER11}. This instrument covers the entire wavelength range from $0.3$ to $2.5\,\mu$m in three spectral arms (UVB arm from the ultraviolet to the $B$ band; VIS arm in the visual regime; NIR arm in the near-infrared regime), at medium resolution ($R\sim3300$ to $18200$, depending on the slit width) simultaneously. This broad wavelength regime gives the opportunity to study the correction of several absorption bands of different species simultaneously (e.g. H$_2$O, O$_2$, CO$_2$).

We have reduced the entire publically available ESO archive data from October 2009 to March 2013, leading to a comprehensive data set taken under various observing conditions, since \xshoot{} is frequently used. We used the ESO standard pipeline in version V2.0.0 under Reflex V2.3 on our cluster\footnote{12 x Intel Xeon X5650@2.67GHz/ 12GB RAM per node}.

For studying the influence of the atmospheric conditions and instrumental set-ups
on the quality of telluric absorption correction (see
Sect.~\ref{sec:results}), we have taken all 1D spectra of TSS without flux
calibration until March 2013. In total,
there are 4218 NIR-arm and 3823 VIS-arm spectra. UVB-arm spectra were not
considered due to the lack of molecular absorption features\footnote{Note
that the ozone absorption by the Huggins and Chappuis bands (see
Paper~I) is usually taken into account by the extinction
correction \citep[see e.g.][]{PAT11}.}. Due to occasional failures of the
automatic \xshoot{} pipeline, the number of obtained spectra is lower than the
number of exposures in the archive. Since the obtained set of reduced data was large enough for our purpose, we did not attempt to re-run the pipeline manually (possibly using tuned reduction parameters) for the cases for which the automatic approach failed.

Two observations of a $\gamma$-ray burst and a Seyfert 2 galaxy (\#1 and 2 in Tables~\ref{tab:tac_comp_dataset1} and \ref{tab:tac_comp_dataset2}) were used to investigate the quality of the telluric absorption correction obtained with \mf{} when applied to data with low $S/N$. Both observations show low count numbers (462 and 241 ADU, respectively) and were also taken at significant airmass, which increases the effect of atmospheric absorption.

For the comparison with the classical method described in Sect.~\ref{sec:comparison}, we have selected science observations of four different astrophysical objects (\#3 through 6 in Tables~\ref{tab:tac_comp_dataset1} and \ref{tab:tac_comp_dataset2}) in conjunction with their corresponding TSS observations manually from our \xshoot{} test data set in order to perform a comparison between the telluric absorption correction achieved
with \mf{} and the classical method related to TSS. The science test data sets were selected to cover different astrophysical objects and $S/N$ traced by the counts in ADU (see Table~\ref{tab:tac_comp_dataset1}) to also estimate the limits of the application of both methods. The TSS spectra were selected to coincide best with the airmass and the date of the corresponding science observation (see Table~\ref{tab:tac_comp_dataset2}) and were usually reduced with the same flat field. As the majority of the telluric absorption features arise in the infrared regime, we restrict this investigation to NIR-arm data.

\begin{table*}
\caption[]{\xshoot{} NIR-arm data set used for the detailed tests (see also Table~\ref{tab:tac_comp_dataset1}).}
\label{tab:tac_comp_dataset2}
\centering
\vspace{5pt}
\begin{tabular}{l | l l l | l l l | c}
\hline\hline
\noalign{\smallskip}
& \multicolumn{3}{c}{Science observations}       & \multicolumn{3}{c}{Telluric standard stars } & \\
 \#$^{a}$   & object               &  airmass  & obs. date $t_{\rm SCI}$      & star & airmass & obs. date $t_{\rm TSS}$ &  $\Delta t_{\rm obs}$$^{b}$ [min]\\
\noalign{\smallskip}
\hline
\noalign{\smallskip}
1 & PKS 1934-63     &  1.69  & 2011-07-01T01:57:23  & -- & -- &  --   & -- \\
2 & GRB 130606A   &  1.715 & 2013-06-07T04:09:16 & -- & -- & -- & -- \\
\noalign{\smallskip}
\hline
\noalign{\smallskip}
3& V921-Sco       &  1.109 & 2010-03-10T09:01:02  & Hip084409 & 1.058 &  2010-03-10T09:42:39   & 42 \\
4& NGC5812       &  1.06  & 2010-03-13T08:57:03  & Hip073345 & 1.128 &  2010-03-13T08:48:11    & 9 \\
5& HE 1331-0247    &  1.075 & 2010-03-24T06:10:55  & Hip085195 & 1.152 &  2010-03-24T07:54:29   & 104 \\
6& IC1266         &  1.222 & 2010-07-05T05:54:38  & Hip085885 & 1.248 &  2010-07-05T06:07:41   & 13 \\

\noalign{\smallskip}
\hline
\end{tabular}
\tablefoot{\\
\tablefoottext{a}{Objects \#1 and \#2 are data with low $S/N$ used for tests on the reliability of \mf{} when applied to such data. \#3 through \#6 are used for a detailed comparison with the classical method. \\}
\tablefoottext{b}{approximate time between TSS and science target observations ($\Delta t_{\rm obs}=\mid\!t_{\rm SCI}-t_{\rm TSS}\!\mid$)}\\
}
\end{table*}

\section{Quality of telluric absorption correction}\label{sec:results}

In the following, we evaluate the quality of telluric absorption correction
with \mf{} by means of a large sample of \xshoot{} TSS spectra (see
Sect.~\ref{sec:data}). We mainly focus on the NIR arm, where the telluric
absorption correction is most crucial. We also complement the discussion with
results from the VIS arm. The \mf{} test set-up for the data set and the
quality indicators used are discussed in Sect.~\ref{sec:approach}. The results
are shown in Sects.~\ref{sec:outliers} to \ref{sec:instru}. The effect of
changing the fitting ranges and \mf{} input parameters is discussed for an
example spectrum in Sects.~\ref{sec:ranges} and \ref{sec:inputpar},
respectively. The influence of differences in the input water vapour profile is
described in Sect.~\ref{sec:profiles}. Finally, we investigate the effect
of low $S/N$ data on the fitting quality in Sect.~\ref{sec:low_sn}.

\subsection{Test set-up and figures of merit}\label{sec:approach}

\begin{table*}
\caption[]{\Mf{} parameter set-up for the telluric absorption correction
evaluation of NIR-arm \xshoot{} TSS spectra}
\label{tab:setup_nir}
\centering
\vspace{5pt}
\begin{tabular}{l l l}
\hline\hline
\noalign{\smallskip}
Parameter$^\mathrm{a}$ & Value & Short description \\
\noalign{\smallskip}
\hline
\noalign{\smallskip}
{\sc wlgtomicron} & $10^{-3}$ & factor to convert the input wavelength units
into $\mu$m \\
{\sc vac\_air} & air & wavelengths in vacuum or air \\
{\sc ftol} & $10^{-2}$ & relative $\chi^2$ convergence criterion \\
{\sc xtol} & $10^{-2}$ & relative parameter convergence criterion \\
{\sc list\_molec} & O$_2$ CO$_2$ H$_2$O CH$_4$ CO & list of molecules to be included in the
model \\
{\sc fit\_molec} & 0 0 1 0 0 & flags for if each molecule is to be fit (1=yes,
0=no) \\
{\sc relcol} & 1. 1.05 1. 1. 1. & relative molecular column densities
(1~=~input profile) \\
{\sc fit\_cont} & 1 & flag for polynomial fit of the continuum (1=yes, 0=no) \\
{\sc cont\_n} & 1 & degree of polynomial for the continuum fit \\
{\sc cont\_const} & $10^{4}$ & initial constant term for the continuum fit \\
{\sc fit\_wlc} & 1 & flag for refinement of wavelength solution (1=yes,
0=no) \\
{\sc wlc\_n} & 0 & degree of Chebyshev polynomial for refined wavelength \\
& & solution \\
{\sc wlc\_const} & 0. & initial constant term of the Chebyshev polynomial \\
{\sc fit\_res\_box} & 1 & flag for resolution fit using a boxcar filter (1=yes,
0=no) \\
{\sc relres\_box} & 0.75 & initial value for FWHM of boxcar relative to slit
width \\
& & ($\geq 0$ and $\leq 2$) \\
{\sc kernmode} & 0 & kernel mode (0=independent Gaussian and Lorentzian, \\
& & 1=Voigt profile) \\
{\sc fit\_res\_gauss} & 1 & flag for resolution fit using a Gaussian filter
(1=yes, 0=no) \\
{\sc res\_gauss} & 1. & initial value for FWHM of Gaussian (in pixels) \\
{\sc fit\_res\_lorentz} & 0 & flag for resolution fit using a Lorentzian filter
(1=yes, 0=no) \\
{\sc res\_lorentz} & 0. & initial value for FWHM of Lorentzian (in pixels) \\
{\sc kernfac} & 3. & kernel size in units of FWHM \\
{\sc varkern} & 1 & flag for selecting a constant (=0) or a variable kernel
(=1) \\
{\sc pixsc} & 0.20 & pixel scale in arcsec \\
\noalign{\smallskip}
\hline
\end{tabular}
\tablefoot{\\
\tablefoottext{a}{Only parameters required for the fitting procedure and
specific to the \xshoot{} data are listed} (see the \mf{} User Manual for more
details}
\end{table*}

\begin{table}
\caption[]{Wavelength ranges (vacuum) for fitting NIR- and VIS-arm \xshoot{}
TSS spectra}
\label{tab:ranges}
\centering
\vspace{5pt}
\begin{tabular}{c c c c l}
\hline\hline
\noalign{\smallskip}
Arm & No. & $\lambda_\mathrm{min}$ [$\mu$m] & $\lambda_\mathrm{max}$ [$\mu$m] &
Main molecule \\
\noalign{\smallskip}
\hline
\noalign{\smallskip}
NIR & 1 & 1.120 & 1.130 & H$_2$O \\
NIR & 2 & 1.470 & 1.480 & H$_2$O \\
NIR & 3 & 1.800 & 1.810 & H$_2$O \\
NIR & 4 & 2.060 & 2.070 & CO$_2$ \\
NIR & 5 & 2.350 & 2.360 & CH$_4$ \\
\noalign{\smallskip}
\hline
\noalign{\smallskip}
VIS & 1 & 0.686 & 0.694 & O$_2$ \\
VIS & 2 & 0.759 & 0.770 & O$_2$ \\
VIS & 3 & 0.930 & 0.945 & H$_2$O \\
\noalign{\smallskip}
\hline
\end{tabular}
\end{table}

\begin{figure}
\centering
\includegraphics[width=9.0cm,clip=true]{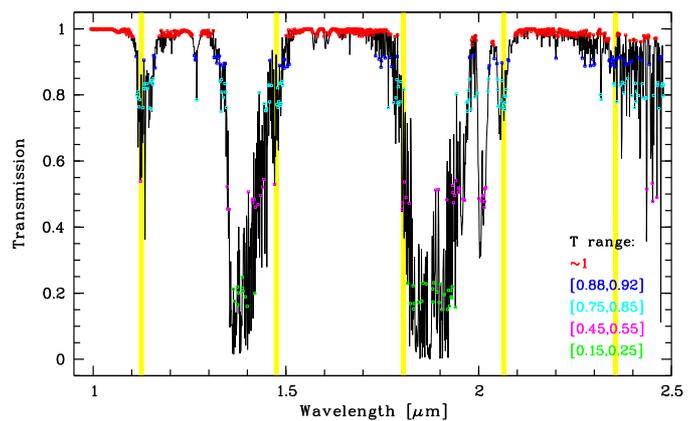}
\caption[]{Reference \xshoot{} NIR-arm model transmission spectrum
binned in 1\,nm stepsfor a mean amount of precipitable water vapour (PWV) of
3.1\,mm and an airmass of 1. The bins marked by red symbols are
classified as continuum. The transmission of these bins ranges from 0.95 to
1.00. The mean value is 0.99. The selected bins are used to derive the
systematic offsets (residuals) in the telluric absorption corrected spectra.
Bins that belong to four specific transmission ranges corresponding to
$T = 0.9$ (blue), 0.8 (cyan), 0.5 (magenta), and 0.2 (green) are indicated as
well. The figure also shows the fitting ranges (yellow bars) that were used by
\mf{} (see also Table~\ref{tab:ranges}).}
\label{fig:method_nir}
\end{figure}

\begin{figure}
\centering
\includegraphics[width=9.0cm,clip=true]{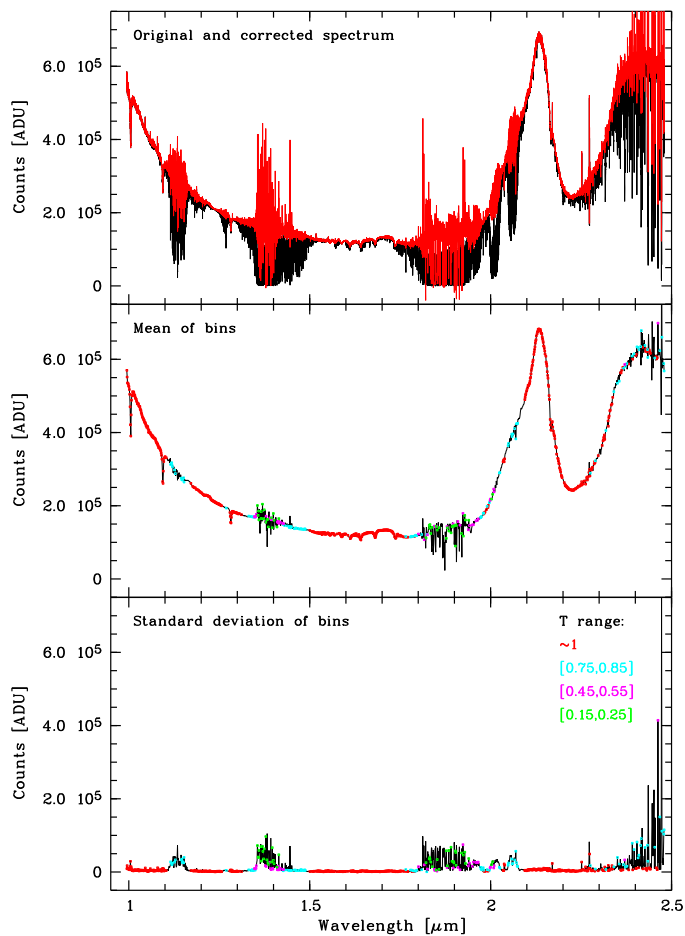}
\caption[]{Quality of telluric absorption correction for an example TSS
spectrum taken with the NIR arm of the \xshoot{} spectrograph. The star was
observed with a 1.2'' slit at an airmass of 1.32. The seeing was 0.76'' and the
PWV (as derived from the fitting) was 1.46\,mm. {\em Upper panel:} The telluric
absorption corrected spectrum (red) is shown in comparison with the original
spectrum (black). {\em Middle and lower panels:} Mean counts and standard
deviation in ADU for a grid of 1\,nm bins for the telluric absorption corrected
example spectrum. The two subfigures also highlight the pixels for the
continuum interpolation (red) and the different transmission ranges for the
quality analysis of the telluric absorption correction. For more details, see
Fig.~\ref{fig:method_nir}.}
\label{fig:taceval_nir_1000}
\end{figure}

\begin{figure}
\centering
\includegraphics[width=9.0cm,clip=true]{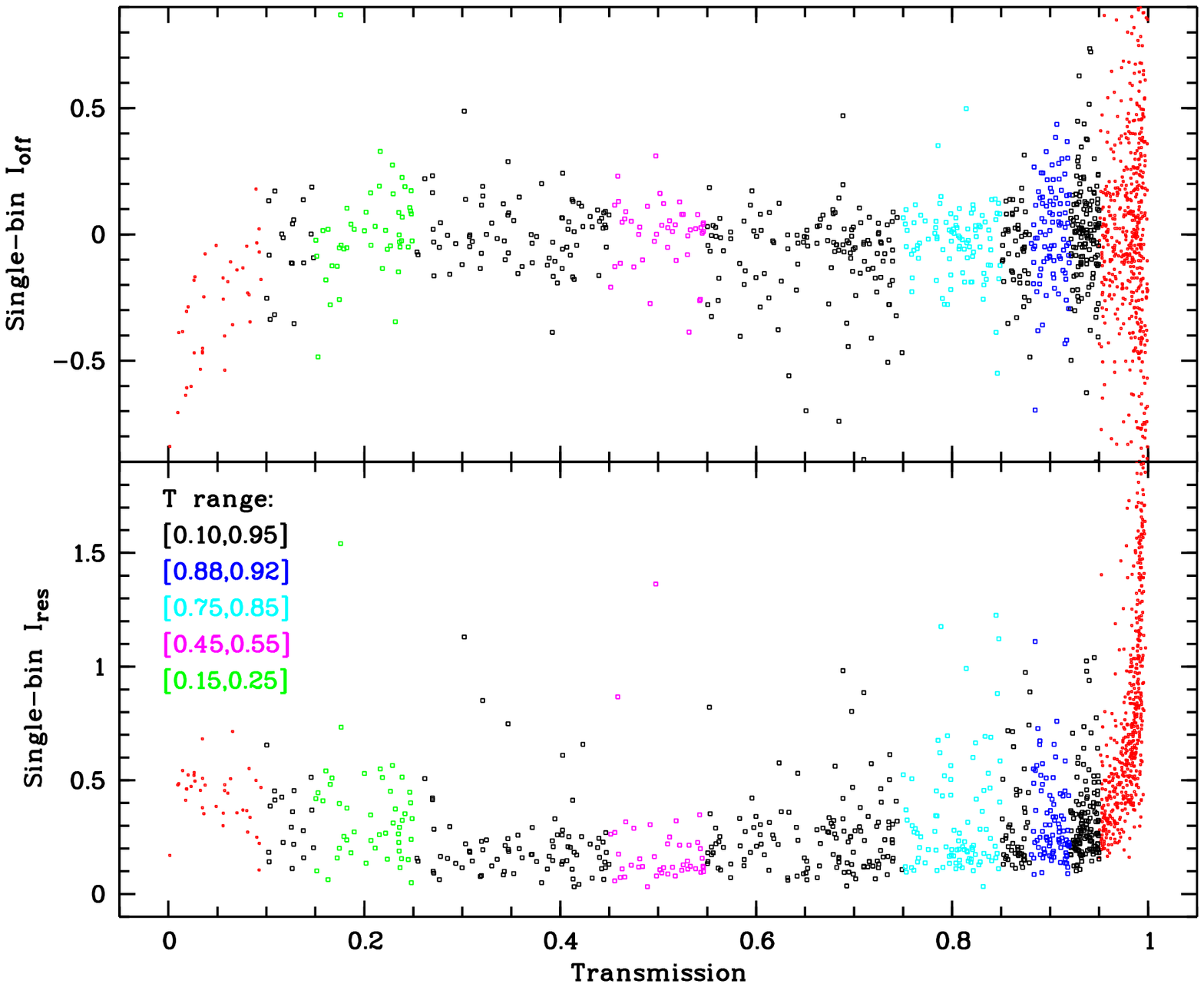}
\caption[]{Indicators $I_\mathrm{off}$ for systematic offsets
({\em upper panel}) and $I_\mathrm{res}$ for small-scale residuals
({\em lower panel}) of the telluric absorption correction for single 1\,nm
bins of the \xshoot{} NIR-arm spectrum shown in
Fig.~\ref{fig:taceval_nir_1000}. Both quantities are given as a function of
transmission. Bins belonging to the different transmission ranges used in the
discussion are marked by different colours (see legend and
Fig.~\ref{fig:method_nir}). Small red symbols identify bins which are excluded
from further analysis due to either very low transmission or very weak line
strength.}
\label{fig:tacevalbin}
\end{figure}

The performance of \mf{} was tested with a fixed input parameter set. This
approach is appropriate to estimate statistically the quality of the
correction on a large set of data. Nevertheless, further quality improvements
could be obtained by adjusting the fitting parameters for each individual
spectrum.

For the NIR arm, the applied set-up is shown in Table~\ref{tab:setup_nir} (see
also Paper~I). For the wavelength range from 1 to 2.5\,$\mu$m, the
model-relevant molecules are O$_2$, CO$_2$, H$_2$O, CH$_4$, and CO. For the VIS
arm, it is sufficient to consider O$_2$ and H$_2$O. Only water vapour is fitted
in spectra of both arms, since the concentration variations and the impact on
the \xshoot{} data of the other species are expected to be small. The
equatorial standard atmospheric profile that we use is already more than a
decade old (prepared by J. J. Remedios 2001; see \citealt{SEI10}). As the
global CO$_2$ concentration increases with time \citep{WMO12}, the input CO$_2$
column was multiplied by 1.05 to be representative of the \xshoot{} archival
data.

A linear fit ({\sc cont\_n}~=~1) was performed to correct the continua of the
spectra in each of the fitting ranges (see below). The initial continuum factor
{\sc cont\_const} was set to $10^4$ to be in the order of the typical count
level in ADU of \xshoot{} data without flux calibration. For the wavelength
grid correction, which is required to handle calibration uncertainties and
an inaccurate centring of the target in the slit, only a constant shift was
allowed ({\sc wlc\_n}~=~0) (see Sect.~\ref{sec:inputpar} for a discussion). For
the instrumental profile, a combination of a boxcar and Gaussian was assumed. A
possible Lorentzian was not considered, since a study of the shape of the line
profiles did not reveal significant Lorentzian wings. The width of the initial
boxcar was chosen to be 75\% of the slit width ({\sc relres\_box}), which
should be close to the real value for the different slits (as confirmed by tests).
An exception is the 5'' slit, which is, however, rarely used for observations
of science targets. For the Gaussian, a reasonable initial FWHM of 1~pixel
({\sc res\_gauss}) and a kernel size of 3~FWHM ({\sc kernfac}) were given.
Since echelle spectra are fitted, the kernel FWHM was selected to be
proportional to the wavelength ({\sc varkern}~=~1). The profile-related input
parameters refer to the central wavelengths of 1.74\,$\mu$m and 0.78\,$\mu$m
for the NIR and VIS arm, respectively. For the 214 spectra (5.1\%) taken with a
$K$-blocking filter \citep[see][]{VER11}, the corresponding wavelength is
1.55\,$\mu$m.

As shown by Table~\ref{tab:ranges} and Fig.~\ref{fig:method_nir}, the fitting
of the atmospheric transmission model in the NIR arm was restricted to five
narrow 10\,nm wide fitting ranges (or inclusion regions; cf. Paper~I).
They cover only about 3\% of the entire wavelength range, which ensures that
the fitting time is reasonable (typically 1 to 2\,min) and the continuum fit
with a low order polynomial is accurate enough. Moreover, only a relatively
small fraction of the NIR-arm wavelength range is suitable for the fits, since
a good fit requires a wide range of transmission values. The windows are
sufficient to derive the amount of atmospheric water vapour (CO$_2$ and CH$_4$
are not fitted), the wavelength shift, and the instrumental profile. Since
the last two properties have to be determined for the entire spectrum,
Ranges~4 and 5, which do not show significant H$_2$O absorption, are also
important for a good coverage of the full wavelength range (see also
Sects.~\ref{sec:ranges} and \ref{sec:inputpar}). Moreover, our optimised set of
fitting ranges avoids absorption features of typical TSS.

The VIS arm is much less affected by molecular absorption. The only
prominent bands are the A and B molecular oxygen bands and the water vapour
band at 0.94\,$\mu$m. Therefore, fitting ranges were only defined in these
three bands as indicated by Table~\ref{tab:ranges}. The H$_2$O-related
range avoids the wavelengths of potentially strong Paschen lines, which
are present in spectra of hot TSS.

Running \mf{} for the input parameters listed in Table~\ref{tab:setup_nir} and
the fitting ranges shown in Table~\ref{tab:ranges} results in the best-fit
parameters for each sample exposure (see Sect.~\ref{sec:data}). These are
then used to provide a telluric absorption corrected spectrum for the full
wavelength range (see Fig.~\ref{fig:taceval_nir_1000}). The quality of the
fit in the pre-defined fitting ranges can be evaluated by considering the
RMS of the residuals, which is provided by \mf{}. However, the quality of
the telluric absorption correction must be studied over the whole spectral
range with respect to the quality of the corrected object spectrum and
therefore requires a different analysis. For this purpose, we have defined the
figures of merit $I_\mathrm{off}$ and $I_\mathrm{res}$. The former measures the
continuum-normalised difference between the original and telluric absorption
corrected spectrum, relative to a locally-averaged telluric absorption
strength. The latter traces the continuum-normalised standard deviation of the
residuals of the correction, relative to a locally-averaged telluric
absorption strength. Therefore, $I_\mathrm{off}$ and $I_\mathrm{res}$ are
indicators of large-scale systematic offsets in the telluric absorption
corrected spectra and small-scale (or high-frequency) variations of the
residuals, respectively. These quality indicators were calculated in the
following way:
\begin{itemize}
\item First, each telluric absorption corrected NIR- and VIS-arm spectrum was
divided into 1\,nm and 0.5\,nm wide bins, respectively (corresponding to about
17 and 25~pixels, i.e. a few resolution elements; cf. Sect.~\ref{sec:instru}).
\item The mean and the standard deviation were calculated for each of these
bins (see Fig.~\ref{fig:taceval_nir_1000}).
\item The bins for which the average transmission model in
Fig.~\ref{fig:method_nir} shows no or only minor absorption were selected as
continuum nodes. Even though this selection includes weak outer wings of some
bands, this is not critical. Lines with a depth of about 1\% can be well
corrected (see Fig.~\ref{fig:taceval_nir_1000}), especially with respect to the
much stronger lines for which the quality of the correction is evaluated.
\item The continuum bins were used to interpolate the object continuum at the
positions of bins with significant absorption.
\item The interpolated continuum intensity for each bin was then subtracted
from the corresponding measured mean. This results in an estimate of systematic
offsets in the telluric absorption corrected spectra.
\item To be independent of the absolute flux, the offsets and standard
deviations of each bin were divided by the interpolated continuum mean values.
\item To make the resulting values independent of the wavelength- and
time-dependent transmission $T$, the relative offsets and standard deviations
were divided by $1 - T$, i.e. the average amount of absorption for each bin
(see Fig.~\ref{fig:tacevalbin}).
\item Five transmission ranges were defined for a detailed analysis of the
telluric absorption correction. Apart from a wide range from 0.1 to 0.95,
ranges centred around 0.9, 0.8, 0.5, and 0.2 were defined (see
Figs.~\ref{fig:method_nir} to \ref{fig:tacevalbin}). We did not consider
$T$ close to 1 and 0, since the related results are mainly noise driven and are
therefore not suited to evaluate \mf{} (see also Sects.~\ref{sec:outliers} and
\ref{sec:obscond}).
\item The bins that belong to each of the five ranges were determined by using
the best-fit transmission spectra of the individual sample spectra. Thus, the
bin assignment depends on the airmass, amount of atmospheric water vapour, and
spectral resolution.
\item Finally, to obtain the figures of merit and to reduce the effect of
outliers, we took the median of the relative offset and standard deviation of
the bins for each transmission range as shown in Fig.~\ref{fig:tacevalbin}.
\end{itemize}
The resulting quality indicators $I_\mathrm{off}$ and $I_\mathrm{res}$ are used in
the subsequent analysis. If the transmission range is not specified explicitly,
the results for the wide range are given. Note that $I_\mathrm{off}$ and
$I_\mathrm{res}$ have to be multiplied by $1 - T$ to provide the systematic
offsets of the correction and the intra-bin standard deviations of the
residuals relative to the object continuum. More details on the interpretation
of our figures of merit can be found in Sect.~\ref{sec:obscond}.

\subsection{Outliers}\label{sec:outliers}

\begin{figure}
\centering
\includegraphics[width=9.0cm,clip=true]{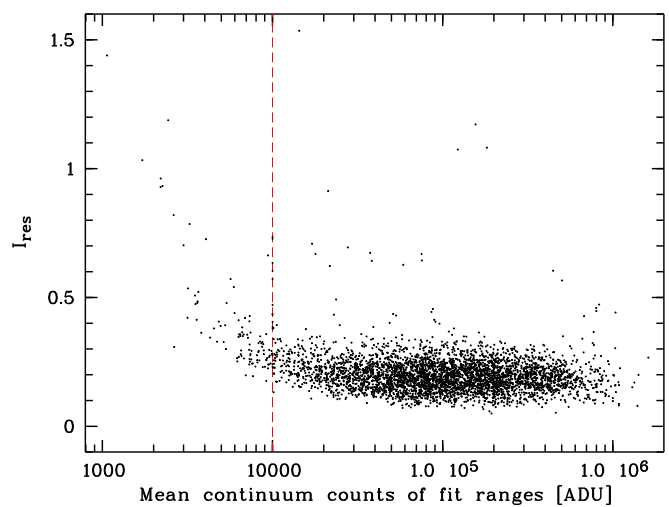}
\caption[]{Small-scale residual indicator $I_\mathrm{res}$ for 1\,nm bins versus
the mean model continuum counts in ADU for the five fitting ranges. The red
dashed line shows the selection criterion for sufficiently high S/N for the
detailed analysis.}
\label{fig:mcont_sig}
\end{figure}

The indicator $I_\mathrm{res}$ measures variations in the count level within
each of the narrow bins of a telluric absorption corrected spectrum. Primarily,
this traces the small-scale quality of the telluric absorption correction.
However, random noise, defects in the spectra, sky subtraction residuals, and
spectral features of the observed object can also cause an increase of
$I_\mathrm{res}$. This is demonstrated for random noise in
Fig.~\ref{fig:mcont_sig}, which shows a clear increase of the scatter for lower
mean counts (calculated for the fitted pixels), i.e. decreasing S/N. In
order to avoid difficulties in interpreting the $I_\mathrm{res}$ sample
statistics, spectra with mean counts less than $10^4$\,ADU are excluded
from further analysis. This threshold concerns 120 spectra of the NIR arm (i.e.
2.8\% of the sample) and 36 spectra of the VIS arm (0.9\%). Excluding these
data does not mean that their corrected spectra have a bad quality. For
example, the mean $I_\mathrm{off}$ for NIR-arm spectra with mean counts
between $10^3$ and $10^4$\,ADU is 0.000 ($\sigma = 0.053$), i.e. there are no
systematic continuum offsets on average. For a discussion of the quality
of the telluric absorption correction for low-S/N spectra, see
Sect.~\ref{sec:low_sn}.

The output files of the \xshoot{} pipeline provide bad pixel masks, which
can be used by \mf{} to exclude critical pixels from the fitting procedure.
Sometimes in the NIR arm, it seems that more pixels were rejected by the
pipeline than required. In the case of a very small number of available pixels,
the fit becomes unreliable. While the standard deviation might even decrease,
systematic offsets are expected to become more significant. In addition,
crucial continuum pixels for the interpolation of ranges with strong absorption
bands could be missing, which makes the derivation of $I_\mathrm{off}$ less
reliable. For this reason, we excluded 247 NIR-arm spectra (5.9\%) with less
than half the maximum number of pixels in the fitting ranges.

So far, we have mainly rejected spectra where a proper calculation of the
quality indicators $I_\mathrm{off}$ and $I_\mathrm{res}$ could not be guaranteed.
However, for evaluating the quality of the telluric absorption correction, it
is important to know the fraction of obviously failed fits. For this purpose,
we studied the best-fit FWHM of the instrumental profile (combined boxcar and
Gaussian) and the best-fit shift of the wavelength grid. Interestingly, only
a small number of NIR-arm spectra (no VIS-arm spectra) showed values which were
clearly separated from the general distribution. 28 best-fit model spectra with
a FWHM above 10~pixels (or below 1.5~pixels) or wavelength shifts of more than
2.5~pixels relative to the sample mean could be identified as clear outliers.
14 of these 28 spectra were already rejected by the critera described above.
\Mf{} appears to show a very robust performance, at least for \xshoot{} TSS
spectra.

For further analysis, we excluded all the discussed spectra with suspicious
fits. This results in subsamples of 3837 NIR-arm (91\% of the full sample) and
3787 VIS-arm spectra (99\%).

\subsection{Influence of line transmission and observing conditions}
\label{sec:obscond}

\begin{table}
\caption[]{Sample averages and standard deviations for the
transmission-dependent indicators $I_\mathrm{off}$ and $I_\mathrm{res}$ of the
quality of the telluric absorption correction of NIR-arm \xshoot{} spectra}
\label{tab:nmed_nir}
\centering
\vspace{5pt}
\begin{tabular}{c c c@{\ \ \ }c c@{\ \ \ }c}
\hline\hline
\noalign{\smallskip}
Ref. $T$ & $T$ range & \multicolumn{2}{c}{$I_\mathrm{off}$} &
\multicolumn{2}{c}{$I_\mathrm{res}$} \\
& & mean & $\sigma$ & mean & $\sigma$ \\
\noalign{\smallskip}
\hline
\noalign{\smallskip}
--- & $0.10 - 0.95$ & $-0.009$ & 0.032 & 0.196 & 0.063 \\
0.9 & $0.88 - 0.92$ & $-0.008$ & 0.044 & 0.209 & 0.071 \\
0.8 & $0.75 - 0.85$ & $-0.024$ & 0.046 & 0.184 & 0.068 \\
0.5 & $0.45 - 0.55$ & $-0.004$ & 0.038 & 0.150 & 0.059 \\
0.2 & $0.15 - 0.25$ & $+0.035$ & 0.063 & 0.237 & 0.136 \\
\noalign{\smallskip}
\hline
\end{tabular}
\end{table}

\begin{figure}
\centering
\includegraphics[width=9.0cm,clip=true]{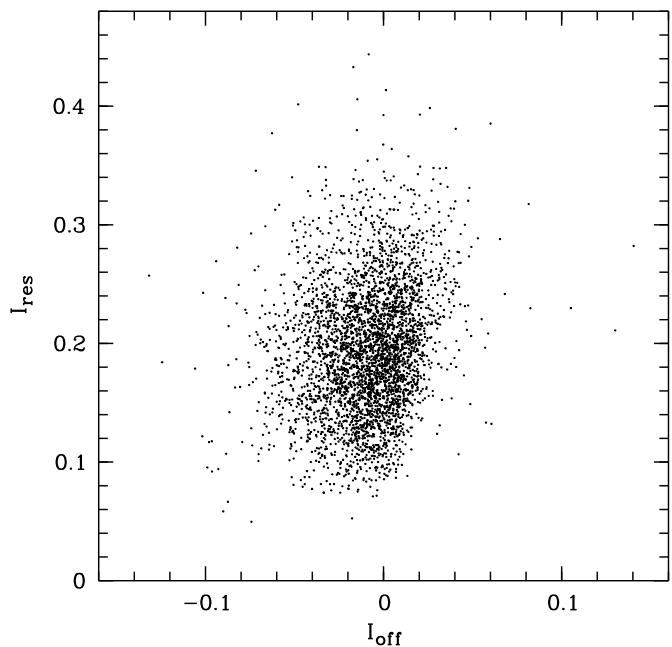}
\caption[]{Indicators $I_\mathrm{off}$ for systematic offsets and $I_\mathrm{res}$
for small-scale residuals of the telluric absorption correction for 1\,nm bins.
Only data points of the filtered subsample are shown.}
\label{fig:nmed_x}
\end{figure}

\begin{figure}
\centering
\includegraphics[width=9.0cm,clip=true]{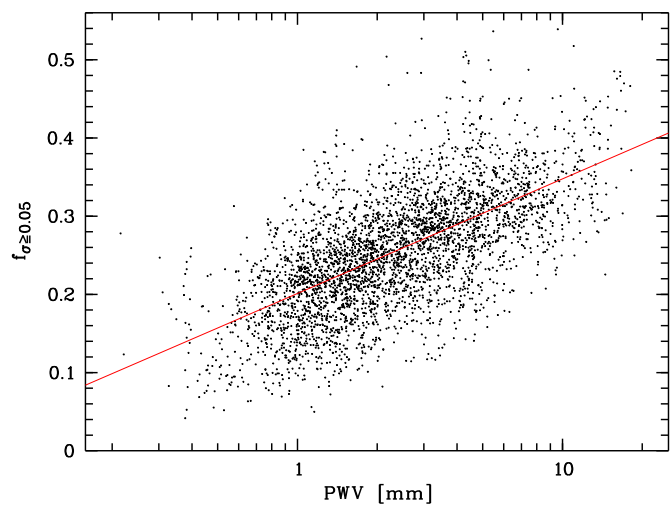}
\caption[]{Fraction of NIR-arm bins with standard deviations of the residuals
of the telluric absorption correction greater than or equal to 5\% of the
object continuum as function of the best-fit PWV in mm. The red regression line
has a slope of 0.15 per dex.}
\label{fig:fbadres_PWV}
\end{figure}

\begin{figure}
\centering
\includegraphics[width=9.0cm,clip=true]{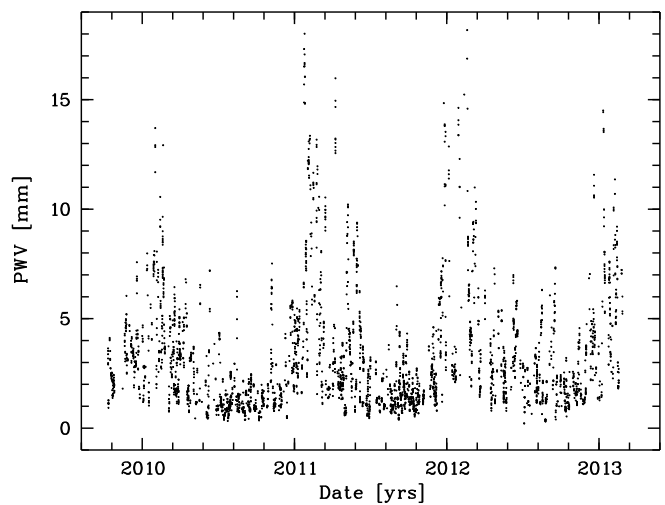}
\caption[]{Best-fit PWV obtained by \mf{} in mm as function of the observing
date in years.}
\label{fig:PWV}
\end{figure}

For the filtered NIR-arm sample (see Sect.~\ref{sec:outliers}),
Table~\ref{tab:nmed_nir} shows the mean values and standard deviations of
$I_\mathrm{off}$ and $I_\mathrm{res}$ for the five transmission ranges listed. In
addition, the individual values for the wide transmission range from 0.1 to
0.95 are plotted in Fig.~\ref{fig:nmed_x}. The latter does not indicate
significant features in the distribution of the data points. $I_\mathrm{off}$
clusters around a value of 0, which means that the telluric absorption
correction does not appear to be affected by systematic offsets. The scatter is
only about 3\% of the line strengths. The mean $I_\mathrm{res}$ is about 0.20
with a scatter of 0.06, i.e. the relative standard deviation of the residuals
of the corrected telluric absorption is about 20\%. This translates into
typical errors of 2\% and 10\% relative to the continuum for spectral
ranges with $T = 0.9$ and 0.5, respectively. The mean $I_\mathrm{off}$ and
$I_\mathrm{res}$ values for the \xshoot{} VIS arm $-0.002$ ($\sigma = 0.048$)
and $0.184$ ($\sigma = 0.059$) are very similar to the NIR-arm results. This
implies that the telluric absorption correction is of good quality in both
\xshoot{} arms. The results are consistent with a typical correction
accuracy of 2\% of the continuum or better for wavelength ranges with
unsaturated telluric lines as reported in Paper~I based on data from different
instruments. Note that individual molecular lines are not resolved in the
\xshoot{} spectra, which lowers and smoothes the measured telluric absorption.

Table~\ref{tab:nmed_nir} lists $I_\mathrm{off}$ and $I_\mathrm{res}$ depending
on the four narrow transmission ranges centred at $T = 0.9$, 0.8, 0.5, and 0.2.
The distributions agree quite well with the results for the wide transmission
range. Consequently, the mixing of transmissions is not crucial for the
resulting quality indicators, at least if $T$ very close to 1 and 0 are not
considered (as for our wide $T$ range). In the case of very high $T$, the
figures of merit are no longer reliable with respect to the quality of the
telluric absorption correction, since random noise, systematic errors
in the reduced spectra, and features of the observed object can have a strong
effect due to the normalisation by $1 - T$ (see Fig.~\ref{fig:tacevalbin}). The
correction quality is probably comparable with the results for intermediate
$T$, which are crucial for the model fit for all $T$. In the case of very low
$T$, the relatively low S/N, possible zeropoint errors, and the strong
variation in $T$ over narrow wavelength ranges can cause high $I_\mathrm{off}$
and $I_\mathrm{res}$. The correction of spectral ranges with $T$ close to 0
is difficult (see Fig.~\ref{fig:taceval_nir_1000}). However, this is usually
not an issue, since the information from the science target also tends to
be very limited. The $I_\mathrm{res}$ values in Table~\ref{tab:nmed_nir}
illustrate the described effects. The minimum of 0.15 is obtained for
intermediate transmissions ($T = 0.5$), whereas the values for $T = 0.9$ and
0.2 are above 0.2.

The rough proportionality of the telluric absorption correction errors
relative to the continuum and $1 - T$ for a wide range of $T$
implies that the overall correction quality of a spectrum is correlated
with properties which affect $T$, i.e. the airmass of the observation and the
column density of the molecular species concerned. This can be understood by
considering that the transmission $T$ is related to the optical depth along the
line of sight $\tau$ for each wavelength $\lambda$ by
\begin{equation}\label{eq:extinction}
T(\lambda) = \mathrm{e}^{-\tau(\lambda)},
\end{equation}
where $\tau$ can be approximated by the product of the optical depth at zenith
$\tau_0$ and the airmass $X$:
\begin{equation}\label{eq:opticaldepth}
\tau(\lambda) \approx \tau_0(\lambda)\,X.
\end{equation}
This works best if the geometrical distributions of the given molecule and air
are similar. Finally, the optical depth $\tau_0$ for molecular absorption by a
single species at wavelength $\lambda$ can be calculated by
\begin{equation}\label{eq:opticaldepth0}
\tau_0(\lambda) = \sigma_\mathrm{abs}(\lambda)
\,\int_{h_0}^{\infty}{n(h)\,\mathrm{d}h},
\end{equation}
where $\sigma_\mathrm{abs}$ is the wavelength-dependent absorption cross section
and the integral corresponds to the column density of the molecule, which is
derived from the density $n$ at heights $h$ above the observer at $h_0$.
Consequently, telluric absorption correction is most difficult if a target is
observed at a large zenith distance and with a high atmospheric water vapour
content. The amount of water vapour is critical, since most prominent
bands in the \xshoot{} wavelength range are caused by this molecule and the
concentration and distribution is highly variable in time and space.

The effect of water vapour on the quality of the telluric absorption correction
is shown in Fig.~\ref{fig:fbadres_PWV}, which displays the fraction of 1\,nm
NIR-arm bins with residuals greater than or equal to 5\% of the corrected
object continuum as function of the amount of precipitable water vapour
(PWV) in mm as derived by \mf{}. For our \xshoot{} NIR-arm TSS sample, the
fitted PWV range from 0.2 to 18.2\,mm with a mean value of 3.1\,mm. The median
is 2.2\,mm. These values are in good agreement with other measurements of the PWV, which is routinely monitored at Cerro Paranal by a stand-alone microwave radiometer in support of science observations \citep[see][and Sect.~\ref{sec:profiles}]{KER14}. There is a clear, nearly linear increase
of the number of bins with significant residuals with increasing PWV in
logarithmic units. A regression analysis indicates that the fraction
$f_{\sigma \ge 0.05}$ grows from about 0.20 at 1\,mm to about 0.35 at 10\,mm. For
comparison, the fraction of bins with a transmission lower than 0.95 of the
mean spectrum plotted in Fig.~\ref{fig:method_nir} is about 0.5. The effect of
the airmass on $f_{\sigma \ge 0.05}$ is smaller than for the PWV, since the
airmass only varies by a factor of about 2. In the VIS-arm range, the
regression line exhibits an increase of $f_{\sigma \ge 0.05}$ from about 0.04 at
1\,mm to 0.07 at 10\,mm. Since strong (water vapour) lines are rare in this
wavelength regime, the fractions are distinctly smaller than in the NIR-arm
range.

The quality of the correction of water vapour bands may roughly depend on
the time of the year. Figure~\ref{fig:PWV} indicates a strong seasonal
dependence of the atmospheric water vapour abundance. In winter, the PWV is
relatively low (mean of 1.9\,mm for meteorological winter), whereas the highest
amounts and a large scatter are found in summer (mean of 5.2\,mm). The strength
of molecular oxygen and carbon dioxide bands can be considered as nearly stable
except for the long-term increase of the CO$_2$ concentration (see
Sect.~\ref{sec:approach}).

\subsection{Influence of resolution}\label{sec:instru}

\begin{table}
\caption[]{Slit-dependent quality of the telluric absorption correction of
NIR-arm \xshoot{} spectra}
\label{tab:slit_nir}
\centering
\vspace{5pt}
\begin{tabular}{@{\ \ }c c c@{\ \ \ }c c@{\ \ \ }c c @{\ \ \ }c@{\ \ }}
\hline\hline
\noalign{\smallskip}
Slit & $N$ & \multicolumn{2}{c}{FWHM$^\mathrm{a}$} &
\multicolumn{2}{c}{$I_\mathrm{off}$} & \multicolumn{2}{c}{$I_\mathrm{res}$} \\
{}[''] & & \multicolumn{2}{c}{[pixels]} & & \\
& & mean & $\sigma$ & mean & $\sigma$ & mean & $\sigma$ \\
\noalign{\smallskip}
\hline
\noalign{\smallskip}
0.4 &  437 & 2.49 & 0.20 & $+0.008$ & 0.059 & 0.239 & 0.069 \\
0.6 &  711 & 3.07 & 0.31 & $-0.002$ & 0.020 & 0.199 & 0.047 \\
0.9 & 1927 & 3.95 & 0.74 & $-0.012$ & 0.027 & 0.193 & 0.064 \\
1.2 &  566 & 4.44 & 1.01 & $-0.015$ & 0.024 & 0.179 & 0.056 \\
1.5 &  148 & 5.03 & 1.62 & $-0.020$ & 0.020 & 0.158 & 0.059 \\
5.0 &   48 & 5.41 & 1.57 & $-0.057$ & 0.025 & 0.156 & 0.066 \\
\noalign{\smallskip}
\hline
\end{tabular}
\tablefoot{\\
\tablefoottext{a}Since the FWHM in pixels depends on the wavelength for
a nearly constant spectral resolution, the FWHM is given for the centre of the
full NIR-arm wavelength range, i.e. 1.74\,$\mu$m (see also
Sect.~\ref{sec:approach}). For spectra taken with a $K$-blocking filter, which
only extend up to 2.1\,$\mu$m, the FWHM was corrected to be also representative
of 1.74\,$\mu$m.}
\end{table}

\begin{figure}
\centering
\includegraphics[width=9.0cm,clip=true]{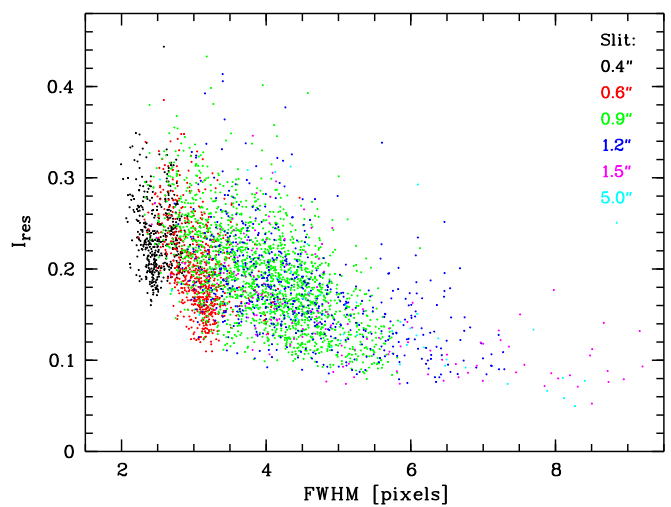}
\caption[]{Small-scale residual indicator $I_\mathrm{res}$ for 1\,nm bins versus
the FWHM of the line profile in pixels. As indicated by the legend, the
different colours correspond to different slit widths.}
\label{fig:FWHM_sig}
\end{figure}

Since the echelle gratings of the \xshoot{} spectrographs are fixed, the slit
width is the only instrumental parameter that affects the line-spread functions
of the resulting spectra. For the NIR and VIS arms, seven different widths
from 0.4'' to 5.0'' can be selected as displayed in
Table~\ref{tab:slit_nir}\footnote{In the case of the VIS arm, the 0.6'' slit is
replaced with 0.7''.}. Apart from the slit selection, the positioning
accuracy of the target in the slit and its change with time contribute to the
resulting line-spread function. Finally, the FWHM of a line is influenced by
the seeing at the time of the observation of the point-like standard star,
especially if the slit is larger than the light profile of the target. For this
reason, Table~\ref{tab:slit_nir} shows an increase of the FWHM (as derived from
the combined best-fit boxcar and Gaussian kernels, see
Sect.~\ref{sec:approach}) as well as its scatter with increasing slit width for
the NIR arm.

The slit-dependent results for the quality indicators $I_\mathrm{off}$ and
$I_\mathrm{res}$ for the wide transmission range are also provided by
Table~\ref{tab:slit_nir}. The systematic offsets appear to indicate a weak
trend from slight overcorrection for the 0.4'' slit to moderate undercorrection
for the 5.0'' slit. However, except for the value for the rarely used 5.0''
slit, the mean offsets can be considered as negligible. Nevertheless, the slit
width seems to contribute to a broadening of the $I_\mathrm{off}$ distribution
of the entire data set (see Fig.~\ref{fig:nmed_x} and
Table~\ref{tab:nmed_nir}).

As indicated by Fig.~\ref{fig:FWHM_sig} and Table~\ref{tab:slit_nir}, the
intra-bin variations of the residuals of the telluric absorption correction
increase with decreasing slit width or FWHM. The mean $I_\mathrm{res}$ values
range from 0.156 for the 5.0'' slit to 0.239 for the 0.4'' slit. For the VIS
arm, the corresponding values are 0.153 and 0.219. This clear dependence
broadens the distribution of $I_\mathrm{res}$ for the entire data set (see
Fig.~\ref{fig:nmed_x} and Table~\ref{tab:nmed_nir}). At first, the trend
can be explained by the expected steepness of the line profiles if a line
comprises only a few pixels (about 2.5~pixels for the 0.4'' slit of the NIR
arm). In this case, small discrepancies between the modelled and the true
profile can cause significant residuals. On the other hand, the scatter in the
NIR/VIS arm is calculated for bins of a width of 1\,nm/0.5\,nm (about 17/25
pixels). If the FWHM is low, more (probably uncorrelated) resolution elements
fit into the bin range. This effect could also augment $I_\mathrm{res}$.

\subsection{Influence of fitting ranges}\label{sec:ranges}

\begin{table}
\caption[]{Influence of fitting ranges on the telluric absorption correction
of the NIR-arm spectrum displayed in Fig.~\ref{fig:taceval_nir_1000}}
\label{tab:tac_ranges_nir}
\centering
\vspace{5pt}
\begin{tabular}{@{\,}c c c c c c c@{\,}}
\hline\hline
\noalign{\smallskip}
Run & Ranges$^\mathrm{a}$ & Rel. & FWHM$^\mathrm{b}$ & PWV & $I_\mathrm{off}$ &
$I_\mathrm{res}$ \\
& & RMS & [pixels] & [mm] & & \\
\noalign{\smallskip}
\hline
\noalign{\smallskip}
1 & 1\,2\,3\,4\,5     & 0.054 & 3.76 & 1.46 & $-0.008$ & 0.235 \\
2 & 1\,--\,--\,--\,-- & 0.030 & 3.19 & 1.68 & $+0.085$ & 0.349 \\
3 & --\,2\,--\,--\,-- & 0.022 & 2.60 & 1.32 & $-0.063$ & 0.246 \\
4 & --\,--\,3\,--\,-- & 0.028 & 3.08 & 1.40 & $-0.041$ & 0.222 \\
5 & 1\,2\,3\,--\,--   & 0.066 & 3.79 & 1.48 & $-0.004$ & 0.238 \\
6 & 1\,--\,--\,--\,5  & 0.035 & 3.29 & 1.56 & $+0.039$ & 0.315 \\
7 & --\,2\,--\,4\,--  & 0.034 & 3.28 & 1.36 & $-0.036$ & 0.235 \\
\noalign{\smallskip}
\hline
\end{tabular}
\tablefoot{\\
\tablefoottext{a}{For the wavelength limits of the different ranges with the
indicated numbers, see Table~\ref{tab:ranges}.\\}
\tablefoottext{b}{The FWHM is given for the centre of the full NIR-arm
wavelength range, i.e. 1.74\,$\mu$m.}}
\end{table}

\begin{table}
\caption[]{Influence of changing fitting ranges in a molecular band on the
telluric absorption correction of the NIR-arm spectrum displayed in
Fig.~\ref{fig:taceval_nir_1000}}
\label{tab:tac_ranges2_nir}
\centering
\vspace{5pt}
\begin{tabular}{@{\,}c@{\ \ }c c c c c c@{\,}}
\hline\hline
\noalign{\smallskip}
Run & Range & Rel. & FWHM$^\mathrm{a}$ & PWV & $I_\mathrm{off}$ &
$I_\mathrm{res}$ \\
& [$\mu$m] & RMS & [pixels] & [mm] & & \\
\noalign{\smallskip}
\hline
\noalign{\smallskip}
2$^\mathrm{b}$ & 1.12$-$1.13 & 0.030 & 3.19 & 1.68 & $+0.085$ & 0.349 \\
2a            & 1.13$-$1.14 & 0.033 & 3.21 & 1.72 & $+0.101$ & 0.399 \\
2b            & 1.14$-$1.15 & 0.039 & 3.03 & 1.42 & $+0.061$ & 0.502 \\
2c            & 1.12$-$1.15 & 0.041 & 3.25 & 1.56 & $+0.100$ & 0.494 \\
\noalign{\smallskip}
\hline
\noalign{\smallskip}
3$^\mathrm{b}$ & 1.47$-$1.48 & 0.022 & 2.60 & 1.32 & $-0.063$ & 0.246 \\
3a            & 1.46$-$1.47 & 0.021 & 2.78 & 1.37 & $-0.044$ & 0.237 \\
3b            & 1.45$-$1.46 & 0.023 & 2.97 & 1.35 & $-0.067$ & 0.256 \\
3c            & 1.44$-$1.45 & 0.023 & 2.97 & 1.35 & $-0.067$ & 0.256 \\
3d            & 1.44$-$1.48 & 0.036 & 2.82 & 1.37 & $-0.045$ & 0.242 \\
\noalign{\smallskip}
\hline
\end{tabular}
\tablefoot{\\
\tablefoottext{a}The FWHM is given for the centre of the full NIR-arm
wavelength range, i.e. 1.74\,$\mu$m.\\
\tablefoottext{b}{same runs as in Table~\ref{tab:tac_ranges_nir}}}
\end{table}

\begin{figure}
\centering
\includegraphics[width=9.0cm,clip=true]{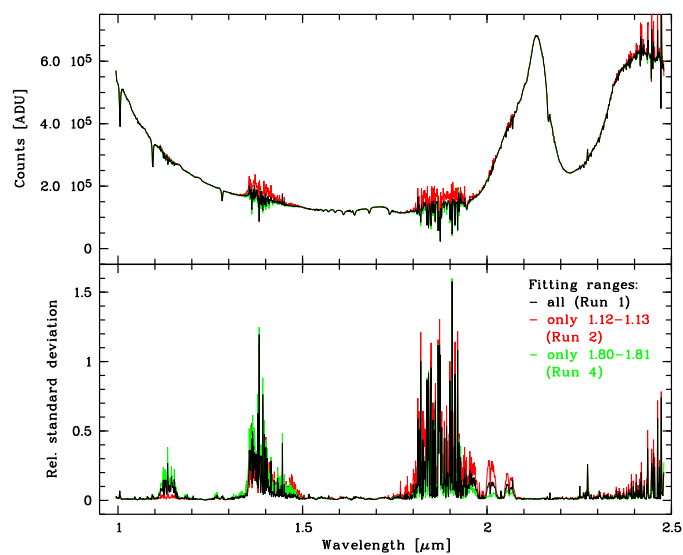}
\caption[]{Mean counts in ADU and standard deviation relative to mean counts
for a grid of 1\,nm bins of the telluric absorption corrected example spectrum
shown in Fig.~\ref{fig:taceval_nir_1000} for three different sets of fitting
ranges. The black spectrum equals the one in Fig.~\ref{fig:taceval_nir_1000}
and corresponds to the standard set-up of windows described in
Table~\ref{tab:ranges}. The red and the green spectra were calculated by only
using a single fitting range (see legend and Table~\ref{tab:tac_ranges_nir}).}
\label{fig:rangetest}
\end{figure}

So far, the discussion has been based on a fixed set of fitting ranges
(see Table~\ref{tab:ranges} and Sect.~\ref{sec:approach}). In the following, we
focus on changes in the quality of the telluric absorption correction when
these ranges are changed. For this purpose, we tested the NIR-arm example
spectrum shown in Fig.~\ref{fig:taceval_nir_1000}, which is characterised by an
airmass of 1.32 and a best-fit PWV of 1.46\,mm. We ran \mf{} for different
subsets of the five standard NIR-arm fitting ranges, which had to include at
least one range dominated by water vapour lines.

Table~\ref{tab:tac_ranges_nir} shows the results of the seven test runs we
performed. The table indicates the selected fitting ranges (see
Table~\ref{tab:ranges} for the numbers), the RMS of the fit residuals
relative to the mean counts, the best-fit FWHM, the best-fit PWV, and the two
quality indicators $I_\mathrm{off}$ and $I_\mathrm{res}$ (see
Sect.~\ref{sec:approach}). The table entries for the different parameters show
a clear difference in the quality of the fit and the telluric absorption
correction depending on the fitting ranges considered. As expected, the RMS of
the fit is reduced if the number of the fitting ranges is decreased. The
resulting FWHM range from 69\% to 101\% of the value of the standard run. For
the PWV, we obtained values from 90\% to 115\%. The largest deviations are
found for runs that were based on only one fitting range. For the quality of
the telluric absorption correction as measured by $I_\mathrm{off}$ and
$I_\mathrm{res}$, there is a similar trend. However, the quality of the
correction also strongly depends on the ranges involved. While the fit only
depending on Range~1 ($1.12$ to $1.13$\,$\mu$m) is by far the worst (Run~2),
the result for Run~4, which is only based on Range~3 ($1.80$ to
$1.81$\,$\mu$m), is remarkably good. This is also illustrated by
Fig.~\ref{fig:rangetest}, which shows the mean values and relative standard
deviations for 1\,nm bins of the resulting spectra of Run~2 and 4 in comparison
with the standard Run~1. Run~2 only led to a good telluric absorption
correction in the fitted range, whereas Run~4 shows a
reasonable correction over the entire wavelength range.

Since the results for the \mf{} runs with single fitting ranges differ
significantly, we performed the model fitting and telluric absorption
correction for single ranges that cover telluric lines in the same bands such
as Runs~2 and 3, but with different wavelength limits. Only ranges where lines
of intermediate strength dominate were selected (see
Fig.~\ref{fig:method_nir}). For ranges within the same H$_2$O band,
Table~\ref{tab:tac_ranges2_nir} reveals similar values for the listed
parameters. This suggests that changing centre and width of a fitting range in
a band (in a reasonable way) has less of an impact than changing the band.

These results imply that modifying the fitting range within the H$_2$O
band at 1.13\,$\mu$m does not significantly improve the quality of the fit. The
PWV, $I_\mathrm{off}$, and $I_\mathrm{res}$ values remain unsatisfying if a
fitting range within this band is not combined with ranges in other molecular
bands. Since the line depths of the different ranges are comparable, this does
not seem to explain the discrepancies. Differences in the best-fit line
shapes and wavelength shifts (up to 1~pixel) for the different \mf{} runs
could imply that the issue is linked to the structure of the \xshoot{}
composite echelle spectra consisting of many orders. At least, line
profiles in the 1.13\,$\mu$m band cover less pixels than those in bands at
longer wavelengths (see Sect.~\ref{sec:instru}). In view of the
uncertainties in the line profile and wavelength calibration, a good telluric
absorption correction over the entire wavelength range (see
Fig.~\ref{fig:rangetest}) requires that all critical molecular absorption bands
are probed by fitting ranges. Therefore, our standard set of fitting windows
(see Table~\ref{tab:ranges}) is well defined, even though only low order
corrections of the different kinds of systematic deviations from the
atmospheric transmission model are possible (see Sect.~\ref{sec:inputpar}). The
best fit is always the result of a compromise, as indicated by the smaller
residuals for individual fitting ranges in the corresponding wavelength regimes
(see Fig.~\ref{fig:rangetest}).

\subsection{Influence of input parameters}\label{sec:inputpar}

\begin{table}
\caption[]{Influence of wavelength grid correction by a Chebyshev polynomial of
order {\sc wlc\_n} on the telluric absorption correction of the spectrum
displayed in Fig.~\ref{fig:taceval_nir_1000}}
\label{tab:tac_grid_nir}
\centering
\vspace{5pt}
\begin{tabular}{c c c c c c}
\hline\hline
\noalign{\smallskip}
{\sc wlc\_n} & Rel. & FWHM & PWV & $I_\mathrm{off}$ & $I_\mathrm{res}$ \\
& RMS & [pixels] & [mm] & & \\
\noalign{\smallskip}
\hline
\noalign{\smallskip}
0 & 0.054 & 3.76 & 1.46 & $-0.008$ & 0.235 \\
1 & 0.126 & 7.25 & 1.51 & $-0.046$ & 0.456 \\
3 & 0.121 & 7.05 & 1.51 & $-0.053$ & 0.459 \\
5 & 0.120 & 6.98 & 1.51 & $-0.050$ & 0.453 \\
\noalign{\smallskip}
\hline
\end{tabular}
\end{table}

\begin{table}
\caption[]{Influence of wavelength grid correction by a Chebyshev polynomial of
order {\sc wlc\_n} on the telluric absorption correction of the spectrum
displayed in Fig.~\ref{fig:taceval_nir_1000} for the fixed best-fit line
profile ({\sc relres\_box}~=~0.560 and {\sc res\_gauss}~=~1.671) as derived
from the standard test run}
\label{tab:tac_grid_noprof_nir}
\centering
\vspace{5pt}
\begin{tabular}{c c c c c c}
\hline\hline
\noalign{\smallskip}
{\sc wlc\_n} & Rel. & FWHM & PWV & $I_\mathrm{off}$ & $I_\mathrm{res}$ \\
& RMS & [pixels] & [mm] & & \\
\noalign{\smallskip}
\hline
\noalign{\smallskip}
0 & 0.054 & 3.75 & 1.47 & $-0.007$ & 0.235 \\
1 & 0.031 & 3.75 & 1.44 & $-0.028$ & 0.195 \\
3 & 0.040 & 3.75 & 1.39 & $-0.042$ & 0.262 \\
5 & 0.031 & 3.75 & 1.42 & $-0.021$ & 0.343 \\
\noalign{\smallskip}
\hline
\end{tabular}
\end{table}

\begin{figure}
\centering
\includegraphics[width=9.0cm,clip=true]{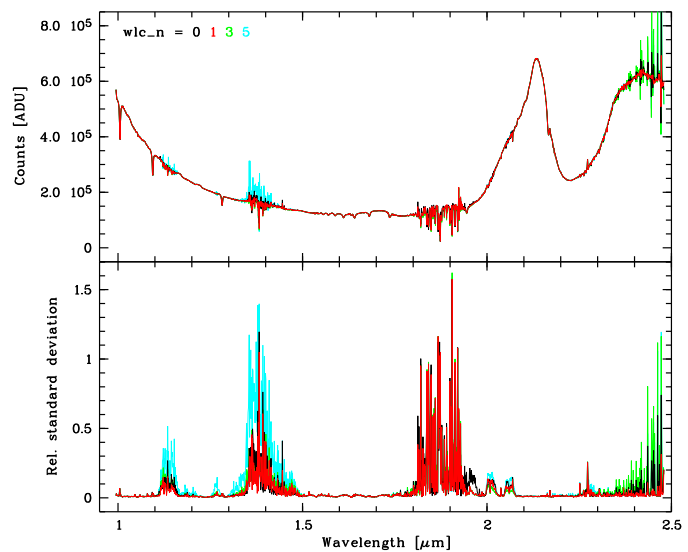}
\caption[]{Mean counts in ADU and standard deviation relative to mean counts
for a grid of 1\,nm bins of the telluric absorption corrected example spectrum
shown in Fig.~\ref{fig:taceval_nir_1000} for different orders of the Chebyshev
polynomial for the wavelength grid correction and fixed line profile (see
legend and Table~\ref{tab:tac_grid_noprof_nir}).}
\label{fig:gridtest}
\end{figure}

Next, we investigate the influence of the input fit parameters on the quality
of the telluric absorption correction. In this respect, our input parameter set
(see Table~\ref{tab:setup_nir}) appears to be a reasonable choice (see
discussion in Sect.~\ref{sec:approach}). However, there could be a critical
restriction of the maximum order of the Chebyshev polynomial for the correction
of the wavelength grid. The selection {\sc wlc\_n}~=~0 only allows a constant
shift of the wavelength grid. As discussed in Sect.~\ref{sec:ranges}, this is
most likely not sufficient to achieve good fits in all the different fitting
ranges at the same time. For this reason, we studied the effect of the
degree of the Chebyshev polynomial on the quality of the fit.
Table~\ref{tab:tac_grid_nir} shows the results of our investigation of the
standard NIR-arm example (see Fig.~\ref{fig:taceval_nir_1000}) for four
different degrees of the polynomial. The result columns are the same as in
Table~\ref{tab:tac_ranges_nir}. The values for the relative RMS, FWHM,
$I_\mathrm{off}$, and $I_\mathrm{res}$ clearly imply that the fits of the runs
with {\sc wlc\_n}~=~1, 3, and 5 failed. The doubling of the FWHM for these
runs indicates an increase of the degrees of freedom by the additional
coefficients of the Chebyshev polynomial caused degeneracies, which led to an
erroneous fit of the instrumental profile. In other words, the fitting
algorithm was not able to find the global $\chi^2$ minimum.

To make the wavelength correction more robust, we performed a second series of
runs with reduced degrees of freedom. For this purpose, we fixed the
properties of the line profile. We took the best-fit results of the standard
run and set {\sc relres\_box}~=~0.560, {\sc res\_gauss}~=~1.671, and the
corresponding fit flags to 0 (cf. Table~\ref{tab:setup_nir}). The results for
the four different degrees of the Chebyshev polynomial are listed in
Table~\ref{tab:tac_grid_noprof_nir}. For the higher {\sc wlc\_n}, the fits in
the five fitting ranges are now better than for the standard run, as the
relative RMS indicate. The PWV values and the related $I_\mathrm{off}$ for
systematic offsets are relatively stable. The small-scale residuals indicator
$I_\mathrm{res}$ is the lowest (0.195) for a linear wavelength correction
function, i.e. {\sc wlc\_n}~=~1. Higher order corrections indicate worse
$I_\mathrm{res}$ (0.343 for {\sc wlc\_n}~=~5). They tend to deteroriate the
telluric absorption correction, despite the fixed line profile. This is also
demonstrated by Fig.~\ref{fig:gridtest}, which shows the mean values and
relative standard deviations for 1\,nm bins of the telluric absorption
corrected spectra for the different runs. For higher degrees of the polynomial,
the quality of the correction seems to be highly wavelength dependent.

These findings suggest that very high {\sc wlc\_n} are risky because of fit
degeneracies. The situation could improve if there were more and/or broader
fitting ranges. However, for the wavelength range covered by \xshoot{},
this approach is not feasible due to the small fraction of wavelengths with
suitably strong absorption lines. Irrespective of these issues, the example has
shown that a combination of two runs (where the second run benefits from the
results of the first run) could significantly improve the quality of the
telluric absorption correction.

For a successful $\chi^2$ minimisation, the number of free parameters should
not be too high. Apart from the wavelength grid correction, the line shape
parameters are prone to $\chi^2$ degeneracies. Line blends by low spectral
resolution and only small numbers of telluric absorption lines by very narrow
fitting ranges can be critical. Fortunately, the \xshoot{} spectra do not seem
to require modelling of possible broad line wings by a Lorentzian kernel
component (see Table~\ref{tab:setup_nir}), which makes the fits more robust.

\subsection{Influence of the input atmospheric profile}\label{sec:profiles}

\begin{figure}
\centering
\includegraphics[width=9.0cm,clip=true]{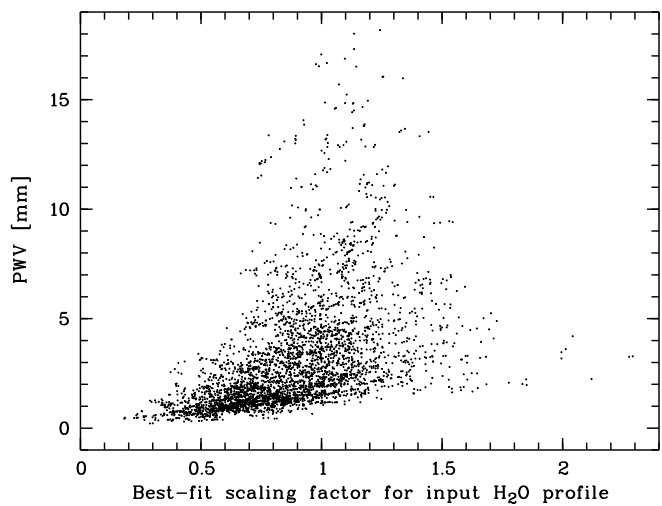}
\caption[]{PWV in mm versus the best-fit scaling factor for the input water
vapour profile (output value for {\sc relcol} parameter; see
Table~\ref{tab:setup_nir}).}
\label{fig:PWV_rH2O}
\end{figure}

Water vapour is an abundant and very variable component of the Earth's
atmosphere (see Fig.~\ref{fig:PWV}). Most telluric absorption in the \xshoot{}
NIR-arm range is caused by this molecule. Therefore,
the quality of the telluric absorption correction strongly depends on a good
fit of the water vapour column density.

The PWV fit corresponds to a scaling of the input water vapour profile. The parameter {\sc relcol} describes the relative scaling with respect to the input profile. Fig.~\ref{fig:PWV_rH2O} shows the final PWV value from the best fit versus the best-fit {\tt relcol} (see Table~\ref{tab:setup_nir}) for the NIR-arm data set selected in Sect.~\ref{sec:outliers}. The mean factor is 0.87 with a scatter of 0.26. This is relatively close to 1, i.e. the case that the PWV of the input profile is the best-fit one. However, for low PWV, the merged input profiles tend to have too much water vapour. For PWV below the median value of 2.2, the mean scaling factor is 0.73, whereas for PWV above the median, a mean factor of 1.01 is obtained. The standard deviation is similar in both cases (0.22 versus 0.23). In view of the significant scatter and the systematic offsets at low PWV, a reliable scaling of the input profiles is indispensable.

To more accurately investigate the effect of the initial atmospheric profiles, we used data of a radiometer, which was installed on Cerro Paranal in October 2011 for water vapour monitoring purposes \citep{KER12a,KER12b}. It is a Low Humidity And Temperature PROfiling microwave radiometer (LHATPRO), manufactured by Radiometer Physics GmbH (RPG\footnote{\tt http://www.radiometer-physics.de/}). The instrument uses several channels across the strong water vapour emission line at 183 GHz, necessary for measuring the low levels of PWV that are common on Cerro Paranal. Details of the radiometer are described in \cite{ROS05}. This radiometer provides continuous direct on-site measurements of the temperature and water vapour content. It also calculates the pressure profile up to a height of about $12$\,km above Cerro Paranal, and the integrated water vapour (IWV), identical to the PWV. We created an additional set of initial atmospheric profiles for \mf{} in the same way as the default combination of a standard atmosphere, GDAS model, and ESO MeteoMonitor described in Paper~I, but replaced the GDAS by the radiometer profiles and skipped the MeteoMonitor data since the latter profiles already contain this information.

From our \xshoot{} data set, we selected 549 telluric standard star observations obtained in Jan, Feb, and Aug-Dec 2012, which correspond to the period of the radiometer data provided to us by ESO. Every spectrum was fitted with \mf{} incorporating both, the GDAS/MeteoMonitor and the LHATPRO based set, respectively. In both cases, we used the closest available profiles. We finally applied a telluric absorption correction based on both methods.

\begin{figure}
\centering
\includegraphics[clip=true]{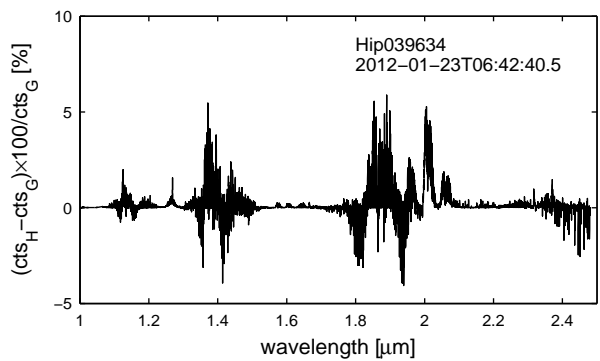}
\caption[]{Relative difference of the telluric absorption corrected spectra of the TSS Hip039634 (IWV\,$=1.5\,$mm), one corrected with a LHATPRO, and one with a GDAS based profile (in per cent). cts$_{\rm H} =$ counts of the LHATPRO based telluric absorption corrected spectrum; cts$_{\rm G} =$ GDAS counterpart.}
\label{fig:gdas_vs_hatro_resi}
\end{figure}

The telluric corrected spectra differ usually by only a few per cent, with larger deviations in the ranges affected by strong atmospheric absorption (see Fig.~\ref{fig:gdas_vs_hatro_resi} for an example).

\begin{figure}
\centering
\includegraphics[clip=true,width=0.48\textwidth]{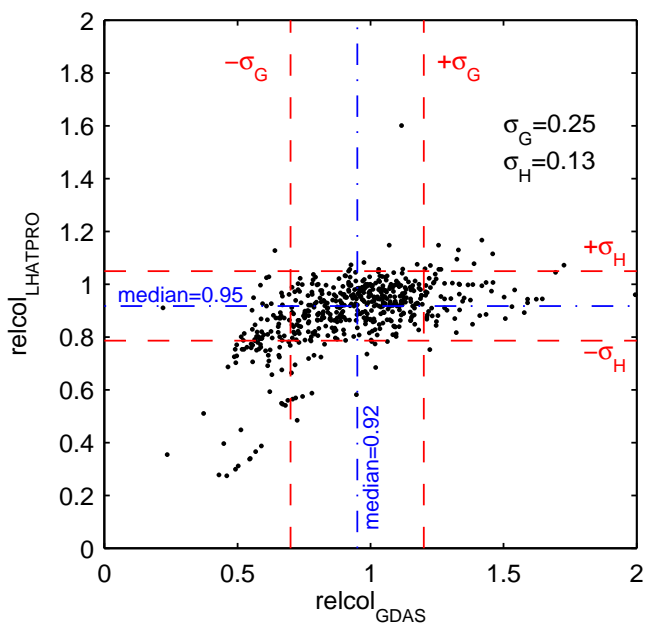}
\caption[]{Comparison of the relative water vapour column scaling factors {\sc relcol} derived with the standard GDAS/MeteoMonitor model and the LHATPRO measurements for 549 TSS spectra. }
\label{fig:gdas_vs_hatro_relcol}
\end{figure}

For a closer look, we also compared the resulting {\sc relcol} values and the water vapour content values, PWV and IWV respectively, for the whole data set. Figure~\ref{fig:gdas_vs_hatro_relcol} gives the comparison of the {\sc relcol} parameters between the GDAS and the LHATPRO based fits. As expected, the LHATPRO based scaling parameters are closer to unity (median value~= 0.95) than the {\sc relcol} parameter derived with the help of the GDAS model (median value~= 0.92). Also the {\sc relcol} scatter $\sigma_{\rm H}=0.13$ for the LHATPRO method is significantly lower than the GDAS based scatter ($\sigma_{\rm G}=0.25$). This means that the atmospheric profile based on the radiometer data requires less scaling than the modelled one. This is expected, since the LHATPRO profiles are direct on-site measurements providing more accurate estimates of the actual atmospheric conditions than the combined GDAS/MeteoMonitor model.


Although the radiometer based initial profiles lead to less scatter in the {\tt relcol} parameter, the quality of the final telluric absorption correction is comparable within a few per cent. This means that due to the adaption with the scaling parameter, an inappropriate initial atmospheric profile also leads to a good telluric absorption correction. We therefore conclude that the underlying fitting algorithm incorporated in \mf{} is highly efficient. However, it appears that the derived PWV value is overestimating the real water vapour content in the case of very dry observing conditions. This indicates inadequate input profiles.

\subsection{Influence of the $S/N$}\label{sec:low_sn}
\begin{figure}
\centering
\includegraphics[clip=true,width=\textwidth/2]{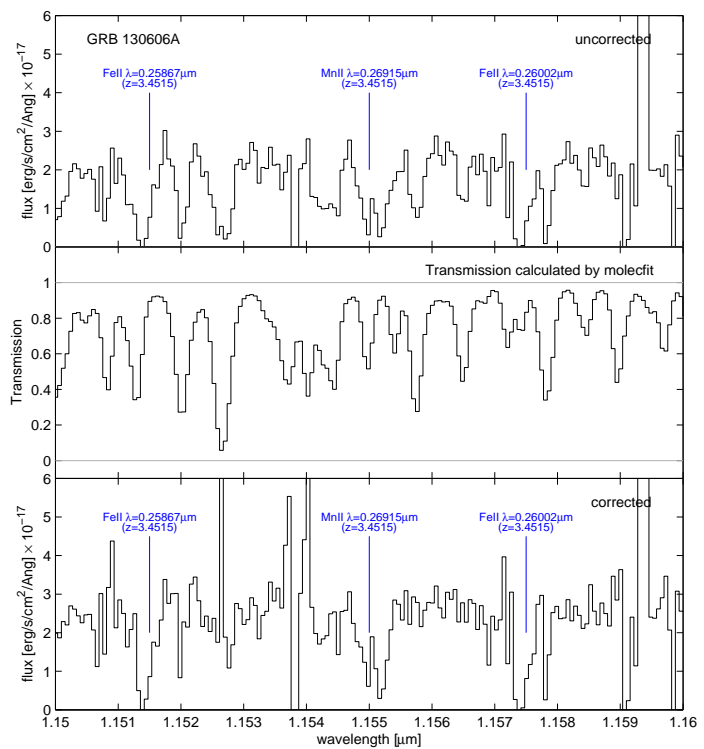}
\caption[]{X-Shooter NIR-arm spectrum of $\gamma$-ray burst GRB 130606A at $z=5.913$. The graphs show the wavelength range from $1.15$ to $1.16\,\mu$m, which is affected by both telluric and some intrinsic absorption lines arising at a redshift $z=3.4515$ \citep{HAR14}. Upper panel: the uncorrected spectrum. Middle panel: transmission spectrum as calculated by \mf{}. Lower panel: corrected spectrum.}
\label{fig:grb}
\end{figure}

In this section, we investigate the performance of \mf{} with respect to low signal data. For this purpose, we used a spectrum of the $\gamma$-ray burst GRB 130606A \citep{XU13, UKW13, CAS13, DEU14} at redshift $z\sim5.913$ \citep{HAR14} taken in June 2013 (Prog.-ID: 091.C-0934; P.I.: Kaper). Fig.~\ref{fig:grb} shows the uncorrected (upper panel), the transmission (middle panel), and the corrected spectrum (lower panel). Although the object flux has a low mean count level of 241 ADU, \mf{} was able to reliably correct the telluric absorption in wide parts.

\begin{figure}
\centering
\includegraphics[clip=true,width=\textwidth/2]{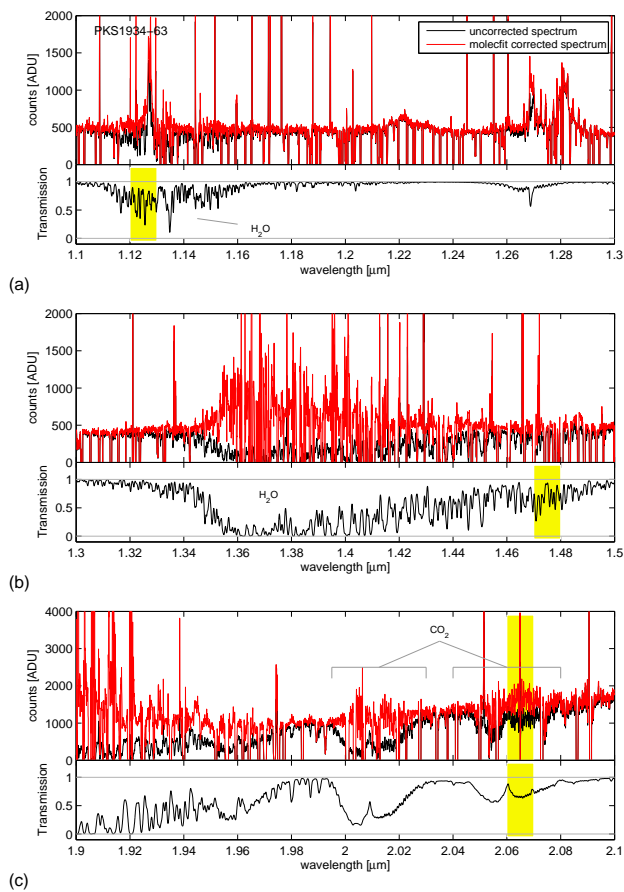}
\caption[]{X-Shooter NIR-arm spectrum of the Seyfert 2 galaxy PKS1934-63 ($z\sim0.183$). The upper panels of (a), (b), and of (c) show the uncorrected and the \mf{} corrected spectrum in different wavelength ranges. This observation has a low mean ADU level of 462 counts. The displayed transmission curve (lower panels) is based on a fit in the standard fitting regions (yellow areas).}
\label{fig:pks}
\end{figure}

Another low $S/N$ example is the spectrum of PKS1934-63, a Seyfert 2 galaxy, taken in July 2011 (Prog.-ID: 087.B-0614; P.I.: Holt). These data contain a comparably low median count level of only 462 ADU (c.f. Table~\ref{tab:tac_comp_dataset1}). Fig.~\ref{fig:pks} shows as an example the three wavelength ranges $1.1$ to $1.3\,\mu$m (upper panel), $1.3$ to $1.5\,\mu$m (middle panel) and $1.9$ to $2.1\,\mu$m (lower panel), respectively. The first two ranges covers the strong water vapour absorption bands between $1.1$ and $1.17\,\mu$m and $1.34$ and $1.52\,\mu$m, respectively. The correction for minor absorption regions is reasonable. The prominent CO$_2$ features between $2.0$ to $2.08\,\mu$m plotted in the lower panel are corrected well.

However, strong absorption bands cause major problems when correcting low ADU data. Fig.~\ref{fig:pks} also shows the strong absorption bands between $1.34$ to $1.52\,\mu$m (panel (a)) and $1.8$ to $1.94\,\mu$m (panel (b)) arising from water vapour and carbon dioxide, respectively. In these regions, the correction leads to large uncertainties, since the incorporated division and the low signal level increase the noise significantly. We therefore conclude that the telluric absorption correction with \mf{} can be critical in such cases. For very low $S/N$ data it might be even impossible to apply \mf{}. In this case, the user might consider observing TSS to derive the correction function.

\section{Comparison with the classical method}\label{sec:comparison}
\subsection{Method}\label{subsec:comparison_method}
\begin{figure*}
\centering
\includegraphics[clip=true,width=\textwidth]{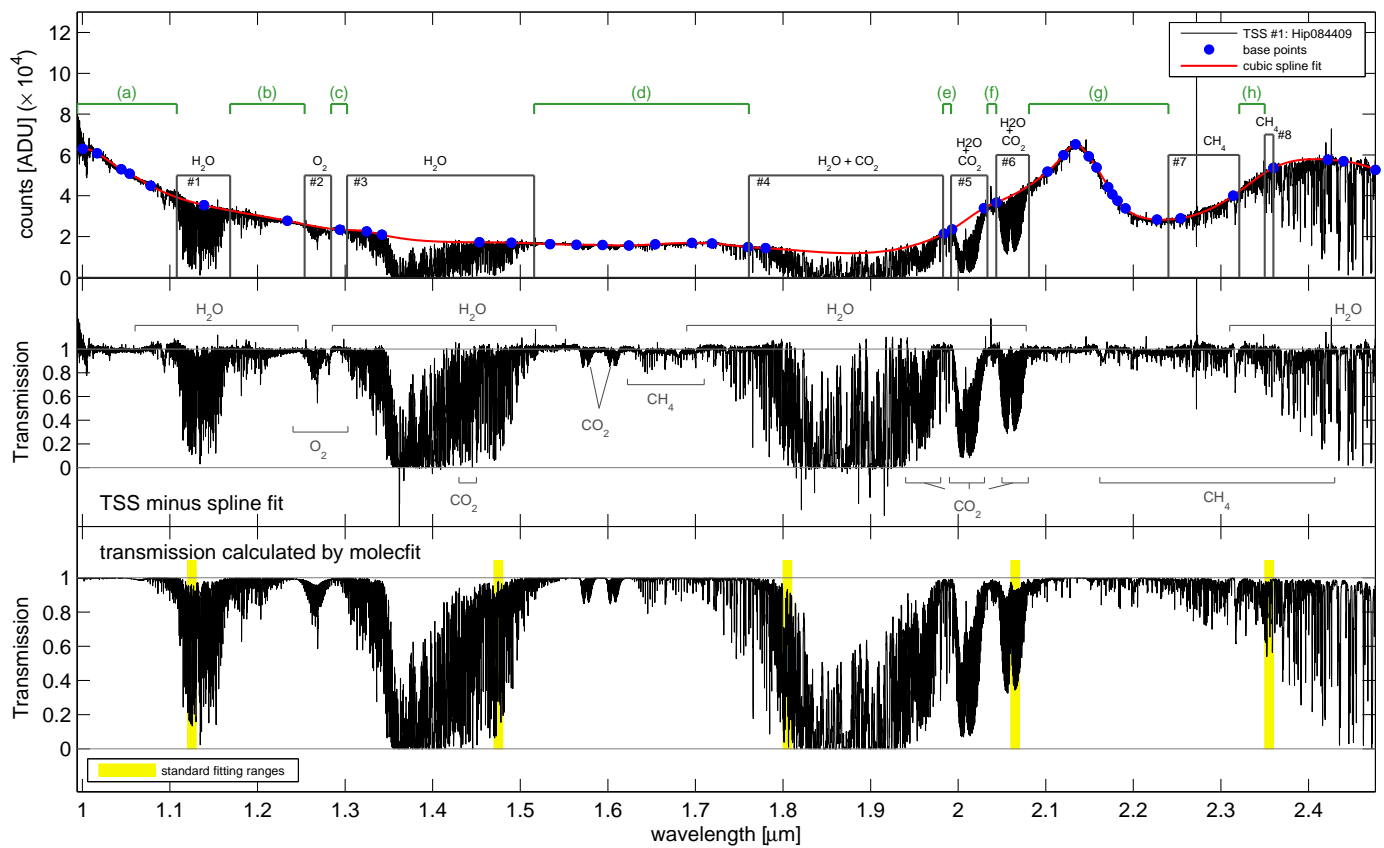}
\caption[]{Upper panel: Telluric standard star spectrum Hip084409, test data set \#3, see Tables\,\ref{tab:tac_comp_dataset1} and \ref{tab:tac_comp_dataset2}) and the fitting method for extracting a transmission curve: A cubic spline fit to base points (blue dots in upper panel) is used to derive the continuum of the telluric standard star. This fit is then used to normalise the spectrum to achieve the corresponding transmission (lower panel). We divided the spectrum into several wavelength regions based on the amount of absorption. Regions \#1 through \#8 are heavily affected by absorption, in contrast to regions (a) through (h) (see Table~\ref{tab:spec_regions1}). Each region was corrected individually with \mf{} and IRAF. Middle panel: The resulting transmission curve including some molecular absorption features. Lower panel: Theoretical transmission curve achieved with \mf{}.}
\label{fig:tsstrans}
\end{figure*}

The classical way of the telluric absorption correction is done with the help of TSS, which are used to derive the transmission of the Earth's atmosphere at the time of the observation. Since TSS are stars with few or well known spectral features, they can be used to obtain an atmospheric transmission after subtracting their continuum and subsequent normalisation. We used the following approach: We fitted a cubic spline to base points selected on positions with or close to transmission $T=1$ in the TSS spectrum in order to determine the continuum. This spline fit is then used to normalise the TSS spectrum resulting in a transmission spectrum (see Fig.~\ref{fig:tsstrans}). We corrected the science spectrum with this transmission using the IRAF task {\tt telluric}\footnote{\url{
http://iraf.net/irafhelp.php?val=telluric&help=Help+Page}}. This task also performs a wavelength shift and a scaling of the input spectrum to achieve the telluric absorption correction.

For the comparison of the quality of the telluric absorption correction between classical TSS and \mf{} methods, we used X-Shooter NIR-arm spectra of four different scientific objects, a B(e) star, an E0 galaxy, a Carbon star, and a planetary nebula (PN), in conjunction with their corresponding telluric standard stars (see Tables~\ref{tab:tac_comp_dataset1} and \ref{tab:tac_comp_dataset2}). We divided the wavelength range covered by the NIR arm into 16 pieces: eight regions with major atmospheric absorption features (labelled with the numbers \#1 through \#8) and another eight regions with minor absorption (labelled with the letters 'a' through 'h', see Fig.~\ref{fig:tsstrans} and Table~\ref{tab:spec_regions1}). 
To optimise the telluric absorption correction, an individual set of fitting parameters was derived for each spectral region with high absorption. Ranges with minor atmospheric absorption were corrected with \mf{} using the standard parameters given in Table~\ref{tab:setup_nir}. To achieve a comparison based on optimised conditions for both methods, the classical method was also applied to each spectral piece individually. In addition, the positions of the base points for the continuum fit were individually chosen for every TSS spectrum to obtain an optimal transmission curve.


\begin{table}
\caption[]{Table of individual regions used for the telluric absorption correction.}
\label{tab:spec_regions1}
\centering
\begin{tabular}{c | c | c }
\hline\hline
\noalign{\smallskip}
region & wavelength  &  included        \\
\#   & range [$\mu$m]    &  molecules \\
\noalign{\smallskip}
\hline
\noalign{\smallskip}
a   & 0.9402 -- 1.1079 & O$_2^{a}$, CO$_2^{a}$, H$_2$O$^{b}$, CH$_4^{a}$, CO$^{a}$\\
b   & 1.1687 -- 1.2537 & O$_2^{a}$, CO$_2^{a}$, H$_2$O$^{b}$, CH$_4^{a}$, CO$^{a}$\\
c   & 1.2800 -- 1.3023 & O$_2^{a}$, CO$_2^{a}$, H$_2$O$^{b}$, CH$_4^{a}$, CO$^{a}$\\
d   & 1.5159 -- 1.7610 & O$_2^{a}$, CO$_2^{a}$, H$_2$O$^{b}$, CH$_4^{a}$, CO$^{a}$\\
e   & 1.9827 -- 1.9918 & O$_2^{a}$, CO$_2^{a}$, H$_2$O$^{b}$, CH$_4^{a}$, CO$^{a}$\\
f   & 2.0333 -- 2.0435 & O$_2^{a}$, CO$_2^{a}$, H$_2$O$^{b}$, CH$_4^{a}$, CO$^{a}$\\
g   & 2.0809 -- 2.2400 & O$_2^{a}$, CO$_2^{a}$, H$_2$O$^{b}$, CH$_4^{a}$, CO$^{a}$\\
h   & 2.3210 -- 2.3500 & O$_2^{a}$, CO$_2^{a}$, H$_2$O$^{b}$, CH$_4^{a}$, CO$^{a}$\\ 
\noalign{\smallskip}
\hline
\noalign{\smallskip}
1   & 1.1079 -- 1.1687  &  H$_2$O$^{b}$\\
2   & 1.2537 -- 1.2800  &  O$_2$$^{b}$ \\
3   & 1.3023 -- 1.5159  &  H$_2$O$^{b}$\\
4   & 1.7610 -- 1.9827  &  H$_2$O$^{b}$, CO$_2$$^{b}$\\
5   & 1.9918 -- 2.0333  &  H$_2$O$^{b}$, CO$_2$$^{b}$\\
6   & 2.0435 -- 2.0809  &  H$_2$O$^{b}$, CO$_2$$^{b}$\\
7   & 2.2400 -- 2.3210  &  CH$_4$$^{b}$\\
8   & 2.3500 -- 2.3600  &  CH$_4$$^{b}$\\
\noalign{\smallskip}
\hline
\end{tabular}
\tablefoot{\\
\tablefoottext{a}{molecular abundance calculated, but not fitted}
\tablefoottext{b}{molecular abundance fitted}}
\end{table}

\subsection{Results}\label{subsec:comparison_results}
\begin{figure}
\centering
\includegraphics[clip=true]{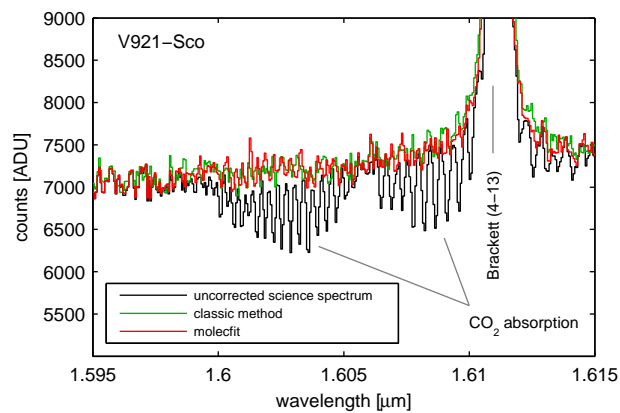}
\caption[]{Comparison of the telluric absorption correction methods: uncorrected object spectrum (black), \mf{} corrected spectrum (red line), and the spectrum corrected with the classical method (green) in a region with minor atmospheric absorption (CO$_2$ band).}
\label{fig:mf_vs_iraf_details1}
\end{figure}

The comparison of the quality of the telluric absorption correction achieved with the classical TSS and \mf{} method reveals that in some regions with minor atmospheric absorption both methods perform similarly, e.g. the weak CO$_2$ bands between $1.6$ and $1.615\,\mu$m are well corrected (see Fig.~\ref{fig:mf_vs_iraf_details1}). There are also some regions with major atmospheric absorption, where both methods achieve very good absorption correction (e.g. redwards of about $2\,\mu$m in Fig.~\ref{fig:mf_vs_iraf_details3}).

However, usually noticeable differences in the quality of the correction are visible. These differences even become critical for regions \#1 through \#8, where major absorption bands affect any ground based observation. In total, we identified three classes of critical problems:
\subsubsection{Object continuum reconstruction}
\begin{figure*}
\centering
\includegraphics[clip=true]{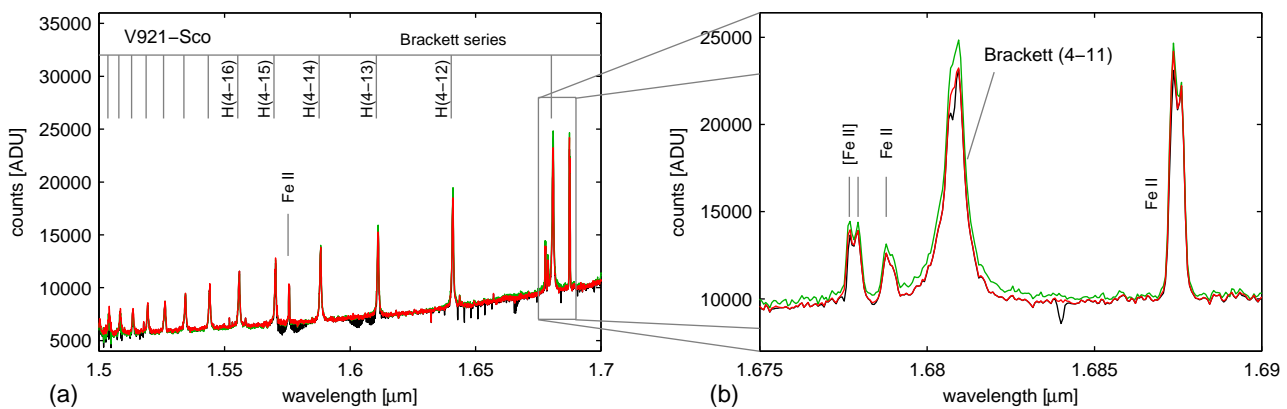}
\caption[]{Panel (a) shows the Brackett series from V921-Sco, a B(e) star. Panel (b) is a zoomed in region with bad continuum correction by the classical method (colour coding as in Fig.~\ref{fig:mf_vs_iraf_details1}).}
\label{fig:mf_vs_iraf_details2}
\end{figure*}

Reconstructing the continuum of the science target in a reliable way is a difficult issue, particularly in regions with broad absorption regions like \#3 and \#4 (see Fig.~\ref{fig:tsstrans} and Table~\ref{tab:spec_regions1}). The quality for the continuum reconstruction with this implementation of the classical method crucially depends on the incorporated interpolation. As such, a fit only can be based on a limited number of base points. Artificial continuum variations are unavoidably introduced in any wavelength range, even with minor or no molecular absorption. Fig.~\ref{fig:mf_vs_iraf_details2} shows an example of a poor continuum reconstruction in a minor absorption region, where the Brackett series and some FeII lines of the B(e) star \citep{KRA12} arise (see Fig.~\ref{fig:mf_vs_iraf_details2}a). The continuum is not well reproduced in the region around the Brackett line (4-11), even though this region is only marginally affected by absorption.
Another example is shown in Fig.~\ref{fig:mf_vs_iraf_details3}, where the spectrum of the E0 galaxy NGC5812 is given in the range of a prominent CO$_2$ absorption region. The continuum shortwards of about $2\,\mu$m derived with the TSS method is too low, leading to an overcorrection of the continuum.

\subsubsection{Line correction/reconstruction}
\begin{figure}
\centering
\includegraphics[clip=true]{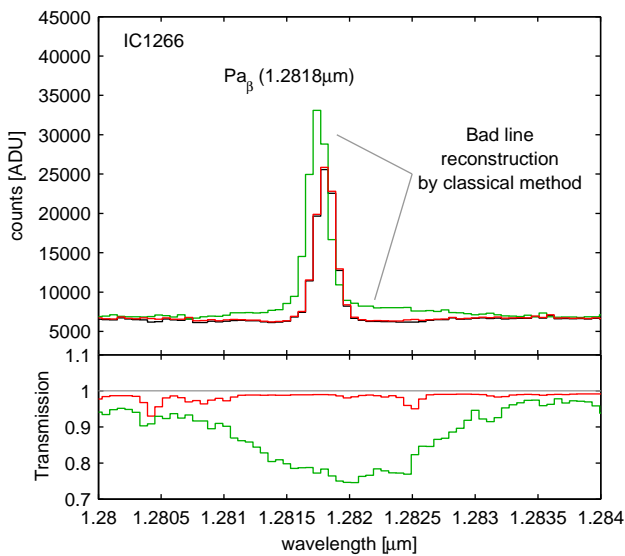}
\caption[]{Pa$_\beta$ line visible in the spectrum of the PN IC1266. Upper panel: Although the line is located in a wavelength range with minor atmospheric absorption, it is not well corrected with the classical method. An intrinsic wavelength shift is probably introduced by the radial velocity of the object  (colour coding as in Fig.~\ref{fig:mf_vs_iraf_details1}). Lower panel: The transmission derived with the classical method (green line) shows a broad dip leading to a significant higher flux and different line shape in the corrected spectrum. The transmission derived with \mf{} (red) does not show this feature.}
\label{fig:mf_vs_iraf_details5}
\end{figure}

Reliable reconstruction of object lines can be a difficult matter for the telluric absorption correction. It becomes critical if a line intrinsic to the TSS affects a scientifically important line in the science spectrum. For example, this applies to hydrogen lines, which is demonstrated by a spectrum of the PN IC1266. Fig.~\ref{fig:mf_vs_iraf_details5} shows the hydrogen Pa$_\beta$ line at $1.2818\,\mu$m, which has significantly higher flux and much broader wings when corrected with the classical method. This is induced by the fitting method to derive the transmission curve, which shows significant differences compared with the one derived with \mf{} (see lower panel of Fig.~\ref{fig:mf_vs_iraf_details5}). In addition, a small wavelength shift is visible in Fig.~\ref{fig:mf_vs_iraf_details5} that is probably caused by a small radial velocity shift affecting the TSS Hip085885.

\begin{figure}
\centering
\includegraphics[clip=true]{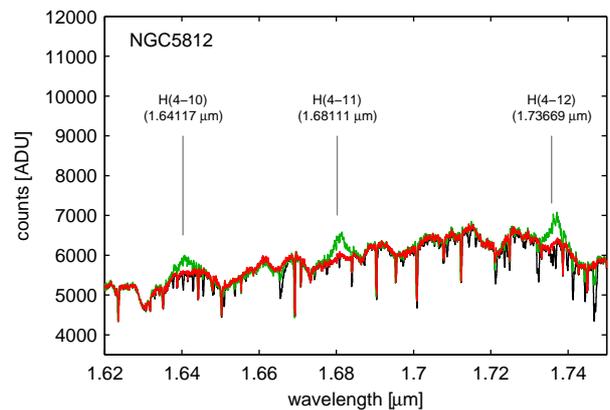}
\caption[]{Example of an overcorrection by spectral features arising from the Brackett lines (H(4-10), H(4-11), and H(4-12)) visible in the TSS. In the \mf{} based correction such features do not occur (colour coding as in Fig.~\ref{fig:mf_vs_iraf_details1}).}
\label{fig:mf_vs_iraf_details7}
\end{figure}

Another example is given in Fig.~\ref{fig:mf_vs_iraf_details7}, which shows a spectrum of the E0 galaxy NGC5812 with significant overcorrection by the classical method arising from TSS Brackett lines. As \mf{} incorporates a purely theoretical transmission curve, such intrinsic stellar features do not occur.

\begin{figure}
\centering
\includegraphics[clip=true]{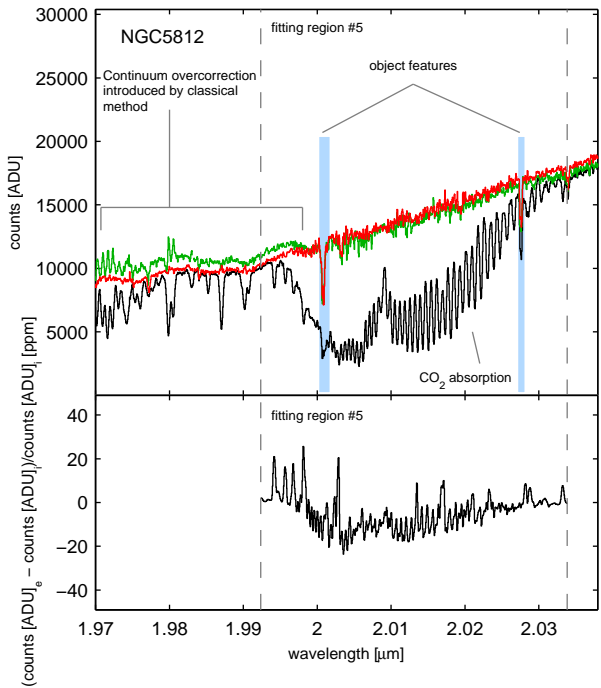}
\caption[]{Upper panel: Example for poor continuum reproduction by the classical method (colour coding as in Fig.~\ref{fig:mf_vs_iraf_details1}). Although the prominent CO$_2$ absorption feature is well corrected by both methods, significant variations in the continuum are visible in the spectrum corrected by the classical method. Intrinsic object lines redwards of $2\,\mu$m are well reconstructed by both methods. Lower panel: relative residuals of the corrected spectra when object lines are excluded from the fit in region \#5 (blue areas in upper panel).}
\label{fig:mf_vs_iraf_details3}
\end{figure}

One of the possible drawbacks of using \mf{} could be the sensitivity of the fit to intrinsic object lines, if they have not been excluded. To investigate this, we used the spectrum of data set \#4, the elliptical galaxy NGC5812, which shows some intrinsic features in region \#5 between $1.995$ and $2.035\,\mu$m (see upper panel of Fig.~\ref{fig:mf_vs_iraf_details3}). This region is dominated by carbon dioxide and minor water vapour absorption. We only used the fitting range \#5 (between $1.9918$ and $2.0333\,\mu$m), where H$_2$O and CO$_2$ were varied. It can be seen that the intrinsic object features remain unchanged (see Fig.~\ref{fig:mf_vs_iraf_details3}). To estimate how sensitive \mf{} is with respect to object lines, we again fitted region \#5, but excluded the object lines from the fit (see blue areas in upper panel of Fig.~\ref{fig:mf_vs_iraf_details3}). We find only marginal differences in the telluric absorption corrected spectra of $\pm20~$ppm. We therefore consider \mf{} to be robust with respect to single object lines, at least if object features do not dominate the corresponding fitting range. However, care should be taken when prominent intrinsic spectral object features or even entire bands are expected at the same wavelength as strong telluric absorption lines. These features might be visible e.g. in low mass stars, molecular clouds, or planetary atmospheres. Prominent features indeed might influence the fit by mimicking differing molecular abundances in the Earth's atmosphere. Hence, the contributions of the astronomical target and the Earth's atmosphere cannot be disentangled by \mf{}.
In this case, the user is advised to either mask these features, if applicable, or choose different fitting ranges to avoid an unintended removal of object features. Alternatively, if such features are known to be visible in the entire spectrum, \mf{} can be applied to a corresponding TSS and subsequent usage of the resulting transmission spectrum for the correction of the science target.

\subsubsection{Increased noise}
\begin{figure}
\centering
\includegraphics[clip=true]{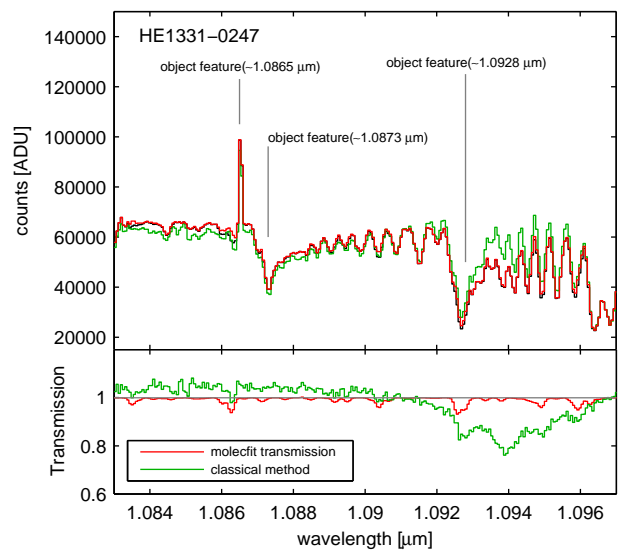}
\caption[]{Upper panel: Detailed zoomed in region with minor atmospheric absorption. Due to variations caused by the spline fit and noise in the TSS spectrum, the classical method introduces continuum variations and noise to the corrected spectrum. All spectral features in the theoretical transmission arise from molecular absorption. Since it is free from random variations, the noise in the corrected spectrum is not increased (colour coding as in Fig.~\ref{fig:mf_vs_iraf_details1}).}
\label{fig:mf_vs_iraf_details8}
\end{figure}

\begin{figure}
\centering
\includegraphics[clip=true]{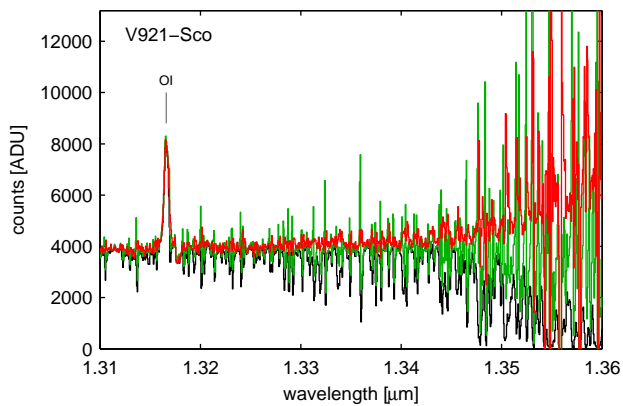}
\caption[]{Close up of the prominent water vapour absorption feature between $1.3$ and $1.5\,\mu$m. Due to the intrinsic noise of the TSS spectrum and the subsequent telluric absorption correction, the noise introduced by the classical method is significantly higher (colour coding as in Fig.~\ref{fig:mf_vs_iraf_details1}).}
\label{fig:mf_vs_iraf_details9}
\end{figure}

\begin{figure}
\centering
\includegraphics[clip=true]{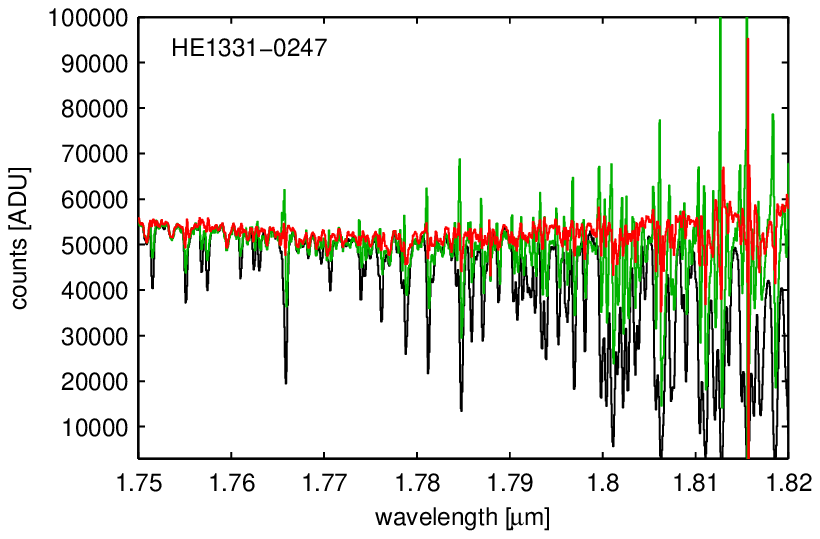}
\caption[]{Same as in Fig.~\ref{fig:mf_vs_iraf_details9} but for the prominent water vapour and carbon dioxide absorption feature between $1.8$ and $2.0\,\mu$m (colour coding as in Fig.~\ref{fig:mf_vs_iraf_details1}).}
\label{fig:mf_vs_iraf_details10}
\end{figure}

The finite $S/N$ of the TSS becomes critical when it is smaller or comparable to the one of the object spectrum. Applying a noisy transmission for the telluric correction derived by means of such a spectrum unavoidably degrades the $S/N$ of the science spectrum. Fig.~\ref{fig:mf_vs_iraf_details8} shows a spectral region with negligible atmospheric absorption, where both methods usually lead to a good reproduction of intrinsic spectral object features (except the variations at $1.094~\mu$m introduced by the TSS transmission). Since the theoretical transmission curve obtained with the radiative transfer code does not show any random noise, \mf{} does not introduce noise-driven features in the science spectrum. However, the classical method does change the noise level. This becomes more and more significant, the more prominent the atmospheric absorption features are. For example, the prominent bands from $1.3$ to $1.5\,\mu$m (H$_2$O) and from $1.8$ to $2.0\,\mu$m (H$_2$O and CO$_2$) lead to very noisy science spectra, when the telluric absorption correction is done with the classical method (see Figs.~\ref{fig:mf_vs_iraf_details9} and \ref{fig:mf_vs_iraf_details10}, respectively). This is particularly important when object features of scientific interest are located there.

\section{Summary and conclusion}\label{sec:summary}
We have developed the software package \mf{} consisting of routines to fit synthetic atmospheric transmission spectra to science data (see Paper~I) and to apply these synthetic spectra as telluric absorption correction to science files. We have extensively tested the software with a large \xshoot{} data set to evaluate the performance of the package with two figures of merit, the offset and the small-scale residual parameters, $I_{\rm off}$ and $I_{\rm res}$, respectively. Moreover, we compared the telluric correction by \mf{} with the classical method based on TSS for several science spectra. In the following, we summarise our findings:
\begin{itemize}
    \item The telluric absorption correction with \mf{} of TSS does not introduce systematic offsets in the corrected spectra. The scatter of $I_{\rm off}$ is about 3\% of the line strength. The relative small-scale residual strength is about 20\% (see Sect.~\ref{sec:obscond} and Fig.~\ref{fig:nmed_x}) for the NIR arm, i.e. the quality of the correction is roughly proportional to the strength of the telluric absorption lines. The VIS arm data show results of similar quality.
    \item The telluric correction shows a dependency on the number of pixels per FWHM. The small-scale residuals increase with decreasing slit width or FWHM (see Sects.~\ref{sec:obscond}, \ref{sec:instru}, and Table~\ref{tab:slit_nir}) if the pixel size is kept constant.
    \item The quality of the fit crucially relies on the selection of the fitting ranges. In particular, all critical molecular absorption bands existing in the wavelength range of the science spectrum should be covered by corresponding regions (see Sect.~\ref{sec:ranges}). On the other hand, a few narrow fitting ranges are sufficient to achieve a good quality of the correction over the entire wavelength range. Spreading these narrow fitting regions over the entire wavelength range helps to improve the fit for the line spread function and the wavelength calibration correction.
    \item \Mf{} offers the possibility to influence the fit by adjustable parameters. The selection of appropriate parameters is crucial for a good result. In particular, the degree of freedom for the fit should the minimised, e.g. the Chebyshev polynomial for the wavelength grid correction has a great influence on the fitting quality. This should be chosen carefully if the fitting ranges are small compared to the entire wavelength range covered by the spectrum as in the case of \xshoot{} (see Sect.~\ref{sec:inputpar}).
    \item Comparisons with atmospheric profiles measured by a microwave radiometer reveal that the fitting algorithm is very robust with respect to variations of the input atmospheric profile. Thus, the initial profile does not have a large effect as long as it does not deviate extremely from the real one. The PWV value determined with \mf{} tends to be too high in the case of very dry observing conditions. However, this does not affect the quality of the telluric corrections.
    \item \Mf{} is also applicable to low $S/N$ data. However, there may be a loss of quality in the telluric correction for very low $S/N$ observations leading to more residuals. Data with extremely low $S/N$ cannot be fitted reliably. In this case, a TSS can alternatively be used for the fit instead of the science spectrum (see Sect.~\ref{sec:low_sn}).
    \item We performed a comparison with the classical method, which is affected by the following problems:
\begin{itemize}
    \item (a) The implementation of the TSS continuum determination can lead to significant continuum changes in the corrected spectrum, e.g. by the limited number of fitting points, and subsequent normalisation of the spectrum.
    \item (b) Intrinsic spectral lines of the TSS can change the flux and shape of object lines or mimic additional spectral features in the corrected object spectrum.
    \item (c) The intrinsic noise of the TSS observation lead to additional noise in the corrected object spectrum.
\end{itemize}
        Whenever such a TSS-related problem appears, it can be expected that \mf{} performs significantly better, as demonstrated in several examples. Moreover, direct fitting of the science spectrum with \mf{} avoids issues related to differences in the atmospheric conditions for the science and TSS observations.
    \item \Mf{} is very robust with respect to single object lines lying in the fitting regions, at least if the regions are not dominated by these features. The telluric correction is only affected marginally in this case. However, care should be taken when molecular bands are expected in the scientific target, which are present in the fitting ranges.
\end{itemize}
The incorporation of synthetic atmospheric transmission spectra based on theoretical calculations provide a promising way to perform the telluric absorption correction. The highly efficient underlying algorithm of \mf{} offers the opportunity to achieve a reasonable and reliable correction without supplementary observations of TSS which are time expensive. In addition, the applicability of \mf{} with standard parameters allows already reasonable results. In addition, optimising these standard parameters can be achieved in much less time than optimising the standard method, making \mf{} a very efficient tool for the telluric absorption correction. We also successfully applied \mf{} to several other ESO instruments covering several wavelength and resolution regimes (see Paper~I). Only a slight adaption of instrument dependent parameters is necessary. Any required information can be added in the parameter file if the FITS header of the files does not contain ESO compliant keywords, which are read by \mf{}. Although the software is delivered with meteorological data for Cerro Paranal, it provides the capability to create atmospheric profiles appropriate also for other observing sites. This flexibility makes \mf{} a general tool for the telluric absorption correction adaptable to various instruments and observing sites.

\begin{acknowledgements}
We thank Sabine M\"ohler (ESO) for the help with the \xshoot{} pipeline and Thomas Kr\"uhler (ESO) for providing us the GRB spectrum. This
project made use of the ESO archive facility. This study was carried out in the
framework of the Austrian ESO In-Kind project funded by BM:wf under contracts
BMWF-10.490/0009-II/10/2009 and BMWF-10.490/0008-II/3/2011. This publication is
also supported by the Austrian Science Fund (FWF): P26130 and by the project IS538003 (Hochschulraumstrukturmittel) provided by the Austrian Ministry for Research (bmwfw).
\end{acknowledgements}

\bibliographystyle{aa}
\bibliography{molecfit_paper_2}

\begin{thebibliography}{20}
\expandafter\ifx\csname natexlab\endcsname\relax\def\natexlab#1{#1}\fi

\bibitem[{{Castro-Tirado} {et~al.}(2013){Castro-Tirado}, {Sanchez-Ramirez},
  {Jelinek}, {Gorosabel}, {Tello}, {Ferrero}, {Lara-Gil}, {Cunniffe},
  {Perez-Ramirez}, {Kubanek}, {Castro Ceron}, {Fernandez-Soto}, {Mottola},
  {Hellmich}, {Fernandez-Munoz}, {Munoz-Martinez}, {Cepa}, \&
  {Alvarez-Iglesias}}]{CAS13}
{Castro-Tirado}, A.~J., {Sanchez-Ramirez}, R., {Jelinek}, M., {et~al.} 2013,
  GRB Coordinates Network, 14790, 1

\bibitem[{{Clough} {et~al.}(2005){Clough}, {Shephard}, {Mlawer}, {Delamere},
  {Iacono}, {Cady-Pereira}, {Boukabara}, \& {Brown}}]{CLO05}
{Clough}, S.~A., {Shephard}, M.~W., {Mlawer}, E.~J., {et~al.} 2005, \jqsrt, 91,
  233

\bibitem[{{de Ugarte Postigo} {et~al.}(2014){de Ugarte Postigo}, {Th{\"o}ne},
  {Rowlinson}, {Garc{\'{\i}}a-Benito}, {Levan}, {Gorosabel}, {Goldoni},
  {Schulze}, {Zafar}, {Wiersema}, {S{\'a}nchez-Ram{\'{\i}}rez}, {Melandri},
  {D'Avanzo}, {Oates}, {D'Elia}, {De Pasquale}, {Kr{\"u}hler}, {van der Horst},
  {Xu}, {Watson}, {Piranomonte}, {Vergani}, {Milvang-Jensen}, {Kaper},
  {Malesani}, {Fynbo}, {Cano}, {Covino}, {Flores}, {Greiss}, {Hammer},
  {Hartoog}, {Hellmich}, {Heuser}, {Hjorth}, {Jakobsson}, {Mottola}, {Sparre},
  {Sollerman}, {Tagliaferri}, {Tanvir}, {Vestergaard}, \& {Wijers}}]{DEU14}
{de Ugarte Postigo}, A., {Th{\"o}ne}, C.~C., {Rowlinson}, A., {et~al.} 2014,
  \aap, 563, A62

\bibitem[{{Hartoog} {et~al.}(2014){Hartoog}, {Malesani}, {Fynbo}, {Goto},
  {Kr{\"u}hler}, {Vreeswijk}, {De Cia}, {Xu}, {M{\o}ller}, {Covino}, {D'Elia},
  {Flores}, {Goldoni}, {Hjorth}, {Jakobsson}, {Krogager}, {Kaper}, {Ledoux},
  {Levan}, {Milvang-Jensen}, {Sollerman}, {Sparre}, {Tagliaferri}, {Tanvir},
  {de Ugarte Postigo}, {Vergani}, {Wiersema}, {Datson}, {Salinas}, {Mikkelsen},
  \& {Aghanim}}]{HAR14}
{Hartoog}, O.~E., {Malesani}, D., {Fynbo}, J.~P.~U., {et~al.} 2014, ArXiv
  e-prints

\bibitem[{{Jones} {et~al.}(2013){Jones}, {Noll}, {Kausch}, {Szyszka}, \&
  {Kimeswenger}}]{JON13}
{Jones}, A., {Noll}, S., {Kausch}, W., {Szyszka}, C., \& {Kimeswenger}, S.
  2013, \aap, 560, A91

\bibitem[{{Kerber} {et~al.}(2014){Kerber}, {Querel}, {Rondanelli}, {Hanuschik},
  {van den Ancker}, {Cuevas}, {Smette}, {Smoker}, {Rose}, \& {Czekala}}]{KER14}
{Kerber}, F., {Querel}, R.~R., {Rondanelli}, R., {et~al.} 2014, \mnras, 439,
  247

\bibitem[{{Kerber} {et~al.}(2012{\natexlab{a}}){Kerber}, {Rose}, {Chac{\'o}n},
  {Cuevas}, {Czekala}, {Hanuschik}, {Momany}, {Navarrete}, {Querel}, {Smette},
  {van den Ancker}, {Cure}, \& {Naylor}}]{KER12a}
{Kerber}, F., {Rose}, T., {Chac{\'o}n}, A., {et~al.} 2012{\natexlab{a}}, in
  Society of Photo-Optical Instrumentation Engineers (SPIE) Conference Series,
  Vol. 8446, Society of Photo-Optical Instrumentation Engineers (SPIE)
  Conference Series

\bibitem[{{Kerber} {et~al.}(2012{\natexlab{b}}){Kerber}, {Rose}, {van den
  Ancker}, \& {Querel}}]{KER12b}
{Kerber}, F., {Rose}, T., {van den Ancker}, M., \& {Querel}, R.~R.
  2012{\natexlab{b}}, The Messenger, 148, 9

\bibitem[{{Kraus} {et~al.}(2012){Kraus}, {Calvet}, {Hartmann}, {Hofmann},
  {Kreplin}, {Monnier}, \& {Weigelt}}]{KRA12}
{Kraus}, S., {Calvet}, N., {Hartmann}, L., {et~al.} 2012, \apj, 752, 11

\bibitem[{{Leinert} {et~al.}(1998){Leinert}, {Bowyer}, {Haikala}, {Hanner},
  {Hauser}, {Levasseur-Regourd}, {Mann}, {Mattila}, {Reach}, {Schlosser},
  {Staude}, {Toller}, {Weiland}, {Weinberg}, \& {Witt}}]{LEI98}
{Leinert}, C., {Bowyer}, S., {Haikala}, L.~K., {et~al.} 1998, \aaps, 127, 1

\bibitem[{{Noll} {et~al.}(2012){Noll}, {Kausch}, {Barden}, {Jones}, {Szyszka},
  {Kimeswenger}, \& {Vinther}}]{NOL12}
{Noll}, S., {Kausch}, W., {Barden}, M., {et~al.} 2012, \aap, 543, A92

\bibitem[{{Patat} {et~al.}(2011){Patat}, {Moehler}, {O'Brien}, {Pompei},
  {Bensby}, {Carraro}, {de Ugarte Postigo}, {Fox}, {Gavignaud}, {James},
  {Korhonen}, {Ledoux}, {Randall}, {Sana}, {Smoker}, {Stefl}, \&
  {Szeifert}}]{PAT11}
{Patat}, F., {Moehler}, S., {O'Brien}, K., {et~al.} 2011, \aap, 527, A91

\bibitem[{{Rose} {et~al.}(2005){Rose}, {Crewell}, {L{\"o}hnert}, \&
  {Simmer}}]{ROS05}
{Rose}, T., {Crewell}, S., {L{\"o}hnert}, U., \& {Simmer}, C. 2005, Atmospheric
  Research, 75, 183

\bibitem[{{Rothman} {et~al.}(2009)}]{ROT09}
{Rothman}, L.~S. {et~al.} 2009, J. Quant. Spectrosc. Radiat. Transfer, 110, 533

\bibitem[{{Seifahrt} {et~al.}(2010){Seifahrt}, {K{\"a}ufl}, {Z{\"a}ngl},
  {Bean}, {Richter}, \& {Siebenmorgen}}]{SEI10}
{Seifahrt}, A., {K{\"a}ufl}, H.~U., {Z{\"a}ngl}, G., {et~al.} 2010, \aap, 524,
  A11

\bibitem[{{Smette} {et~al.}(2014){Smette}, {Sana}, {Noll}, {Horst}, {Kausch},
  {Kimeswenger}, {Szyszka}, {Jones}, {Gallene}, {Vinther}, \&
  {Ballester}}]{SME14}
{Smette}, A., {Sana}, H., {Noll}, S., {et~al.} 2014, paper I, acc. by \aap

\bibitem[{{Ukwatta} {et~al.}(2013){Ukwatta}, {Barthelmy}, {Gehrels}, {Krimm},
  {Malesani}, {Marshall}, {Maselli}, {Melandri}, {Palmer}, \&
  {Stamatikos}}]{UKW13}
{Ukwatta}, T.~N., {Barthelmy}, S.~D., {Gehrels}, N., {et~al.} 2013, GRB
  Coordinates Network, 14781, 1

\bibitem[{{Vernet} {et~al.}(2011){Vernet}, {Dekker}, {D'Odorico}, {Kaper},
  {Kjaergaard}, {Hammer}, {Randich}, {Zerbi}, {Groot}, {Hjorth}, {Guinouard},
  {Navarro}, {Adolfse}, {Albers}, {Amans}, {Andersen}, {Andersen}, {Binetruy},
  {Bristow}, {Castillo}, {Chemla}, {Christensen}, {Conconi}, {Conzelmann},
  {Dam}, {de Caprio}, {de Ugarte Postigo}, {Delabre}, {di Marcantonio},
  {Downing}, {Elswijk}, {Finger}, {Fischer}, {Flores}, {Fran{\c c}ois},
  {Goldoni}, {Guglielmi}, {Haigron}, {Hanenburg}, {Hendriks}, {Horrobin},
  {Horville}, {Jessen}, {Kerber}, {Kern}, {Kiekebusch}, {Kleszcz}, {Klougart},
  {Kragt}, {Larsen}, {Lizon}, {Lucuix}, {Mainieri}, {Manuputy}, {Martayan},
  {Mason}, {Mazzoleni}, {Michaelsen}, {Modigliani}, {Moehler}, {M{\o}ller},
  {Norup S{\o}rensen}, {N{\o}rregaard}, {P{\'e}roux}, {Patat}, {Pena}, {Pragt},
  {Reinero}, {Rigal}, {Riva}, {Roelfsema}, {Royer}, {Sacco}, {Santin},
  {Schoenmaker}, {Spano}, {Sweers}, {Ter Horst}, {Tintori}, {Tromp}, {van
  Dael}, {van der Vliet}, {Venema}, {Vidali}, {Vinther}, {Vola}, {Winters},
  {Wistisen}, {Wulterkens}, \& {Zacchei}}]{VER11}
{Vernet}, J., {Dekker}, H., {D'Odorico}, S., {et~al.} 2011, \aap, 536, A105

\bibitem[{{World Meteorological Organization}(2012)}]{WMO12}
{World Meteorological Organization}. 2012, WMO Greenhouse Gas Bulletin, 8, 1

\bibitem[{{Xu} {et~al.}(2013){Xu}, {Malesani}, {Schulze}, {Fynbo}, {D'Elia},
  {Goldoni}, {Hartoog}, {Hjorth}, {Kaper}, {Kruehler}, {Levan},
  {Milvang-Jensen}, {Tanvir}, \& {Wiersema}}]{XU13}
{Xu}, D., {Malesani}, D., {Schulze}, S., {et~al.} 2013, GRB Coordinates
  Network, 14816, 1

\end{thebibliography}




\end{document}